\documentclass[a4paper]{aa}
\usepackage{graphicx,amsfonts,amssymb,txfonts,natbib,latexsym}
\newcommand{\plotwd}{8.2cm}
\newcommand{\plotwdtwo}{16.4cm}

\titlerunning{Clustering of SZ clusters} 
\title{Clustering of SZ clusters on a past light-cone: acoustic oscillations and constraints on dark energy}

\author{G. H\"utsi\inst{1,2}}

\institute{Max-Planck-Institut f\"ur Astrophysik, Karl-Schwarzschild-Str. 1,
86740 Garching bei M\"unchen, Germany
\and Tartu Observatory, T\~oravere 61602, Estonia}

\offprints{G.H\"utsi , \\ \email{gert@mpa-garching.mpg.de}}
\date{Received / Accepted}
\begin{document}
\abstract {We study the clustering of SZ-selected galaxy clusters on a past light-cone, particularly paying attention to the possibility of constraining properties of dark energy. The prospects of detecting baryonic features in the cluster power spectrum for a wide and shallow survey like PLANCK, and for an SPT-like narrow and deep survey are discussed. It is demonstrated that these future blank sky SZ surveys will have the capability to improve over the recently announced detection of baryonic oscillations based on the SDSS Luminous Red Galaxy (LRG) sample. We carry out parameter estimation using a Fisher matrix approach taking into account the anisotropic nature of the power spectrum due to redshift space and cosmological distortions. The clustering signal which is not too sensitive to systematic uncertainties serves as a valuable piece of information that in combination with other sources of data helps in breaking degeneracies between the cosmological parameters.
}
\maketitle

\section{Introduction}
In the early 1970's it was recognized that acoustic waves in the radiation dominated matter prior to the epoch of recombination of hydrogen in the Universe resulted in the characteristic pattern of maxima and minima in the post-recombination matter power spectrum \citep{1970Ap&SS...7....3S,1970ApJ...162..815P,1978SvA....22..523D}. These acoustic peaks depend on the size of the sound horizon and on the relative phases of the perturbations containing different masses at the moment of recombination. 
In the currently most favourable cosmological models with a Cold Dark Matter (CDM) component dominating signifiantly over the baryonic part, the acoustic features in the matter power spectrum are strongly damped, reaching only $\sim 5 \%$ level for the ``concordance'' model \citep{1999Sci...284.1481B,2003ApJS..148..175S}. For early description of the acoustic oscillations in the context of the CDM models see \citet{1988ApJ...326..539B}.
For a given set of parameters of the Universe ($\Omega_b h^2$, $\Omega_m h^2$) the position of the maxima and minima are fully determined  (e.g. \citealt{1998ApJ...496..605E}). 
Acoustic oscillations also leave their imprint on the angular perturbations of the Cosmic Microwave Background (CMB).   
MAXIMA-1, \footnote{http://cosmology.berkeley.edu/group/cmb/} Boomerang, \footnote{http://cmb.phys.cwru.edu/boomerang/} WMAP, \footnote{http://map.gsfc.nasa.gov} VSA, \footnote{http://www.mrao.cam.ac.uk/telescopes/vsa/} CBI, \footnote{http://www.astro.caltech.edu/~tjp/CBI/} and many other CMB experiments detected the first acoustic peaks in the CMB power spectrum with a high confidence level \citep{2000ApJ...545L...5H,2002ApJ...571..604N,2003ApJS..148....1B,2003MNRAS.341L..23G,2003ApJ...591..556P}. These observations gave very important information about the key parameters of the Universe using the angular scale of acoustic features as rulers and taking into account the ratio of amplitudes of the different peaks. 

The existence of acoustic rulers in the Universe is of enormous importance since they permit us to measure the behavior of the Hubble parameter with redshift and also allow us to establish the distance-redshift relation. This is especially important now when the discovery of \emph{dark energy} (DE) is introducing more questions than the answers it provides. Different areas of observational cosmology (SNe~Ia e.g. SCP ,\footnote{http://panisse.lbl.gov} high-z SN search; \footnote{http://cfa-www.harvard.edu/cfa/oir/Research/supernova/HighZ.html} large scale structure surveys e.g. SDSS, \footnote{http://www.sdss.org/} 2dF \footnote{http://www.aao.gov.au/2df}; CMB experiments) provide evidence that the expansion of our Universe has been proceeding in an accelerated fashion since $z \sim 0.75$ \citep{1998AJ....116.1009R,1999ApJ...517..565P,2003ApJS..148..175S}. Currently there is no physical understanding or even a reliable model for the DE. One of the first tasks to understand the nature of DE will be the measurement of its equation of state parameter, $w = P/\rho$, and its possible evolution with time $w(z)$. When we are equipped with a good standard ruler and are able to measure its angular behavior with redshift, we can obtain very valuable information about $w(z)$. With the CMB data we can determine an angular diameter distance to the last scattering surface with high precision. By combining this information with the measurement of acoustic peaks in the distribution of baryons at lower redshifts $0<z<1 - 2$, we will have unique information about the effective $w$ and may even be able to determine $w(z)$ \citep{1998ApJ...496..605E,2003ApJ...594..665B,2003PhRvD..68f3004H,2003PhRvD..68h3504L,2003ApJ...598..720S}.   

It was obvious since the first publications that acoustic oscillations should also leave their imprint on the large scale structure of the Universe and thus influence e.g. the correlation function and the power spectrum of galaxies. The first successful detection of these features was presented by \citet{astro-ph/0501171} who found traces of acoustic oscillations in the distribution of luminous red galaxies for which they had excellent measurements of angular positions and redshifts obtained by the SDSS collaboration. 

In this paper we discuss the opportunities that will be opened by the planned blank sky deep SZ cluster surveys which will be performed in the coming years. Clusters of galaxies are especially interesting objects for the study of acoustic features in the spatial distribution of objects since it has long been known that \emph{the clustering of clusters is an order of magnitude enhanced in comparison to galaxies} \citep{1983ApJ...270...20B,1984ApJ...284L...9K}. Therefore even with smaller statistics of clusters it is possible to get useful results. 

In the next few years there will be very deep SZ cluster surveys of the restricted regions of the sky performed by several projects, e.g. APEX, \footnote{http://bolo.berkeley.edu/apexsz/} SZA, \footnote{http://astro.uchicago.edu/sza/} AMI, \footnote{http://www.mrao.cam.ac.uk/telescopes/ami/} ACT \footnote{http://www.hep.upenn.edu/act}. Our analysis showed that the volume of these surveys and the number of possible cluster detections will be unfortunately insufficient for the search for the acoustic wiggles in the power spectrum. However, two planned surveys which will be carried out by the PLANCK Surveyor \footnote{http://astro.estec.esa.nl/Planck} spacecraft and the South Pole Telescope \footnote{http://astro.uchicago.edu/spt} (SPT) have very good prospects for the detection of acoustic features. PLANCK will make a shallow blank sky cluster survey permitting one to detect up to $20,000$ rich clusters of galaxies (e.g. \citealt{2004ApJ...613...41M}) with the bulk of objects at $z<0.5$, but will reach distances of $z \sim 0.8$. In contrast the SPT survey will observe deeper, but will cover only $10\%$ of the sky. This survey is expected to detect up to $30,000$ clusters of galaxies \citep{2004ApJ...613...41M}, and many of them will be at significantly higher redshifts compared to the ones observed by PLANCK. Unfortunatelly it is not enough to measure only the SZ flux or brightness of the clusters. In order to measure the equation of state of DE \emph{we need the redshift estimate for each cluster in the sample} which will be hard and time consuming work for many optical, X-ray and possibly radio astronomers. However, when this problem is solved, cosmologists will have a unique sample of clusters of galaxies with good knowledge of their angular position, redshift, and hopefully also mass. In this paper we investigate what limits to the DE equation of state might be obtained when these large experimental efforts are completed. It is obvious that in parallel other ways to measure $w(z)$ will be implemented, but any additional and independent information will be useful. Especially important is that PLANCK and SPT surveys of clusters of galaxies will be performed in any case. Certainly, for many various purposes: (i) study of the redshift distribution of clusters, (ii) study of the properties of the clusters as a population, (iii) search for high-$z$ clusters etc., we always need to estimate redshift. Therefore the information on the power spectrum of clusters of galaxies, acoustic wiggles and the subsequent determination of $w$ is complementary but extremely important part of these surveys. 

The structure of this paper is as follows. In Sec. 2 we describe an analytical model for a cluster power spectrum on a light-cone and calibrate it against the numerical simulations. Sec. 3 discusses the possibility of detecting baryonic oscillations with the forthcoming SZ surveys. In Sec. 4 we carry out parameter forecasting using a Fisher matrix approach and Sec. 5 contains our conclusions.

\section{Light-cone power spectrum of galaxy clusters}
In this section we present the theoretical model for calculating the cluster power spectrum on our past light-cone. In order to assess the accuracy of the theoretical description, we make comparisons with the VIRGO Hubble Volume simulation outputs. We start with a very brief description of the VIRGO simulations and proceed with the calculation of power spectra using light-cone cluster catalogs provided by the VIRGO Consortium \footnote{http://www.mpa-garching.mpg.de/Virgo/}.

\subsection{Cluster power spectra from VIRGO simulations}
We use outputs from the $\Lambda$CDM Hubble Volume simulation that was run in a $3000$ $h^{-1}\,\mathrm{Mpc}$ comoving box with a particle mass of $2.25 \cdot 10^{12} h^{-1}M_{\odot}$. The other simulation parameters were as follows: $\Omega_{\rm m}=0.3$, $\Omega_{\Lambda}=0.7$, $\Omega_{\rm b}=0.04$, $h=0.7$ and $\sigma_8=0.9$ (for further details see e.g. \citealt{2002ApJ...573....7E}). In our study we used $z=0$ snapshot, SphereB and OctantB light-cone catalogs \footnote{For the exact description of these catalogs see \citet{2002ApJ...573....7E}}. The $z=0$ cluster catalog used a friend-of-friend scheme with a linking length $b=0.164$ for cluster identification, while for the light-cone outputs the spherical overdensity method with the overdensity $200$ relative to the critical density was applied. The minimum number of particles per cluster is $30$ and $12$ for the $z=0$ and light-cone catalogs, respectively.

To calculate the power spectrum we follow the direct method of \citet{1994ApJ...426...23F} (FKP), which is shown to be optimal for sufficiently large $k-$modes i.e. $k \gg 1/L$, where $L$ is the typical spatial extent of a survey volume \citep{1998ApJ...499..555T}. Because FFTs are used to achieve significant speedup for Fourier sum calculations, we first have to find the density field on a grid. To this end we use the Triangular Shaped Cloud (TSC) \citep{1988csup.book.....H} mass assignment scheme. Thus our density field is a filtered version of the underlying field, and as shown in \citet{astro-ph/0409240}, the real power spectrum $P$ can be expressed as the following sum over aliases (correct again for the case $k \gg 1/L$):
\\
\begin{eqnarray} \label{eq1}
P_{\rm{raw}}(\mathbf{k}) \simeq \sum_{\mathbf{k'}} S^2(\mathbf{k'}) \sum_{\mathbf{n}\in\mathbb{Z}^3} & &\mathcal{W}^2(\mathbf{k}+2k_{\rm N}\mathbf{n}) P(\mathbf{k}+2k_{\rm N}\mathbf{n}) + 
\nonumber \\
& &\frac{1}{N} \sum_{\mathbf{n}\in\mathbb{Z}^3} \mathcal{W}^2(\mathbf{k}+2k_{\rm N}\mathbf{n}), 
\end{eqnarray}
where the raw power spectrum:
\begin{equation} \label{eq2}
P_{\rm{raw}}(\mathbf{k}) \equiv \langle \vert \delta_{\rm{g}}(\mathbf{k}) \vert ^2 \rangle
\end{equation}
and Fourier transform of the overdensity field on a grid is calculated as usual:
\begin{equation} \label{eq3}
\delta_{\rm{g}}(\mathbf{k})= \frac{1}{N}\sum_{{\rm g}}\left [ n_{\rm g}(\mathbf{r}_{\rm g}))-\bar{n}S(\mathbf{r}_{\rm g})\right ] {\rm e}^{{\rm i}\mathbf{r}_{\rm g} \cdot \mathbf{k}}.
\end{equation}
Here $n_{\rm g}$ is the number density field on a grid without any selection effect corrections, $S$ is a selection function that also incorporates survey geometry (i.e. $S=0$ outside of survey boundaries), $\bar{n}$ is the mean underlying number density and the sum runs over all grid cells. A Fourier transform of the selection function $S(\mathbf{k})$ in Eq. (\ref{eq1}) is normalized so that $S(\mathbf{0})=1$ and the mass assignment window in the case of the TSC scheme can be expressed as:
\begin{equation}\label{eq4} 
\mathcal{W}(\mathbf{k})=\left [ \frac{\prod \limits_{{\rm i}} \sin (\frac{\pi k_{\rm i}}{2k_{\rm N}} )}{\prod \limits_{{\rm i}} \frac{\pi k_{\rm i}}{2k_{\rm N}}} \right ] ^3.
\end{equation}
The second term on the left-hand side of Eq.(\ref{eq1}) is the shot noise contribution and in the case of the TSC filter can be shown to give the following result \citep{astro-ph/0409240}:
\begin{equation} \label{eq5}
{\rm SN} = \frac{1}{N} \prod \limits_{{\rm i}} \left [ 1 - \sin^2 \left (\frac{\pi k_{\rm i}}{2k_{\rm N}} \right ) + \frac{2}{15} \sin^4 \left (\frac{\pi k_{\rm i}}{2k_{\rm N}} \right ) \right ].
\end{equation}
Summarized very briefly, our power spectrum calculation consists of the following steps:
\begin{enumerate}
\item Determination of the selection function $S$ (including survey geometry) and mean underlying number density $\bar{n}$,
\item Calculation of the overdensity field on a grid using the TSC mass assignment scheme and its Fourier transform as given in Eq. (\ref{eq3}), 
\item Subtraction of the shot noise term (Eq. (\ref{eq5})) from the raw power spectrum (Eq. (\ref{eq2})),
\item Isotropization of the shot noise corrected power spectrum, i.e. averaging over $k$-space shells,
\item Application of normalization correction due to selection effects, i.e. dividing by $\sum_{\mathbf{k}} S^2(\mathbf{k})$,
\item Deconvolving the smearing effect of the TSC mass assignment.  
\end{enumerate}
The ``sharpening'' in the last step is done using an iterative method as described in \citet{astro-ph/0409240} with the only difference that here we do not approximate the power spectrum simply with a power law, but also allow for a running of the spectral index, i.e. we approximate it with a parabola in log-log coordinates.

When calculating the power spectrum in the above described way we assumed that the influence of selection/survey geometry effects on the power spectrum are separable. This is not the case for large scales ($k \sim 1/L$) and also if too narrow (i.e. $\Delta k \lesssim 1/L$) power spectrum bins are used. In the following we always make the power spectrum binning broad enough so that the neighboring bins can be safely assumed to be uncorrelated.

The power spectrum error is estimated using the simple ``mode counting'' result of FKP (see also \citealt{1998ApJ...499..555T}):
\begin{equation}\label{eq6}
\frac{\Delta P}{P} = \sqrt{\frac{2}{V_{\rm eff}V_{\rm k}}}, 
\end{equation}
where $V_{\rm k}=4\pi k^2\Delta k/(2\pi)^3$ is the volume of the $k-$space shell and $V_{\rm eff}$ is the effective volume given by:
\begin{equation}\label{eq7}
V_{\rm eff} = \frac{\left[\int W^2(z)\frac{{\rm d}V_{\rm c}}{{\rm d}z}{\rm d}z\right]^2}{\int W^4(z)\left[1+ \frac{1}{\bar{n}(z)P}\right]^2\frac{{\rm d}V_{\rm c}}{{\rm d}z}{\rm d}z}.
\end{equation}
Here ${\rm d}V_{\rm c}$ is a comoving volume element and the weight function:
\begin{equation}\label{eq8}
W(z)\propto \left\{ 
\begin{array}{lll}
\rm{const} & \rm{\quad for\ volume\ weighting}\\
\bar{n}(z) & \rm{\quad for\ number\ weighting}\\
\frac{\bar{n}(z)}{1+\bar{n}(z)P} & \rm{\quad for\ an\ optimal\ FKP\ weighting.}
\end{array} \right.
\end{equation}
\begin{figure}
\centering
\includegraphics[width=\plotwd]
{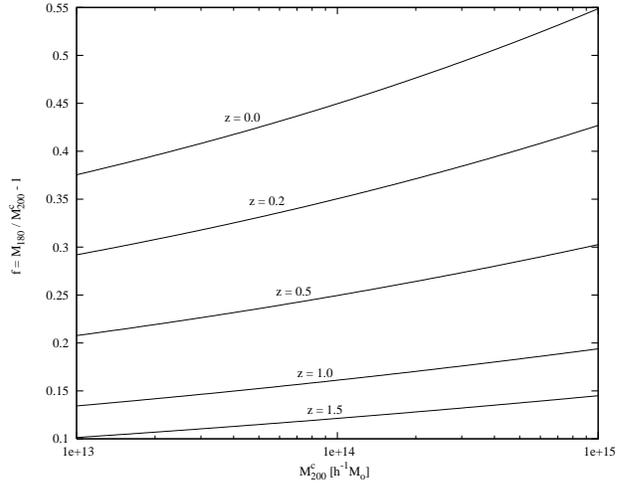}
\caption{Mass conversion from $M^{\rm{C}}_{200}$ to $M_{180}$ for various redshifts.}
\label{fig1}
\end{figure}

The halos of the VIRGO Hubble Volume simulations were identified using the spherical overdensity algorithm with an overdensity of $200$ with respect to the critical density at the identification epoch \citep{2002ApJ...573....7E} (we denote the corresponding mass $M^{\rm{C}}_{200}$), while detailed comparisons suggest that the Press-Schechter type of analytical calculations \citep{1974ApJ...187..425P,1996MNRAS.282..347M,1999MNRAS.308..119S,2001MNRAS.323....1S} provide a good match to simulations if an overdensity $180$ with respect to the background density is used (corresponding mass $M_{180}$) (e.g. \citealt{2001MNRAS.321..372J}). In order to convert from one mass definition to the other we assume that the density profile of clusters is given by the NFW \citep{1997ApJ...490..493N} profile:
\begin{equation}\label{eq9}
\rho(r) = \frac{\rho_{\rm s}}{\left(\frac{r}{r_{\rm s}}\right)\left(1+\frac{r}{r_{\rm s}}\right)^2},
\end{equation}
and the concentration parameter $c=r_{\rm v}/r_{\rm s}$ ($r_{\rm v}$-virial radius) and its evolution as a function of virial mass $M_{\rm v}$ is given as follows \citep{2001MNRAS.321..559B}:
\begin{equation}\label{eq10}
c(M_{\rm v})= \frac{9}{1+z}\left(\frac{M_{\rm v}}{M_{\rm *}(z=0)}\right)^{-0.13}. 
\end{equation}
Here $M_{\rm *}$ is a standard nonlinear mass scale defined through $\sigma(M_{\rm *},z)\equiv\delta_{\rm c}(z)$, where $\sigma^2(M,z)$ is the variance of the linearly evolved density field on the comoving scale corresponding to the mass $M$ at redshift $z$ and $\delta_{\rm c}(z)$ is the spherical collapse threshold e.g. in Einstein-de Sitter model $\delta_{\rm c}(0)=1.686$.

Then the mass within the radius $r$ can be expressed as:
\begin{equation} \label{eq11}
M(<r) = 4\pi\rho_{\rm s}r^3f\left(\frac{r_{\rm s}}{r}\right)=\frac{4\pi r^3}{3}\Delta\Omega_{\rm m}\rho_{\rm c},
\end{equation}
where
\begin{equation}\label{eq12}
f(x) = x^3\left[\ln\left(1+\frac{1}{x}\right)-\frac{1}{1+x}\right],
\end{equation}
$\rho_{\rm c}$ is the critical density and $\Delta$ is the halo overdensity with respect to the background matter density at the epoch of halo identification. 

In order to convert halo mass $M$ corresponding to the overdensity $\Delta$ to the one corresponding to the overdensity $\Delta'$ we proceed as follows:
\begin{enumerate}
\item From Eq. (\ref{eq11}) determine the radius $r$ corresponding to the mass $M$ and overdensity $\Delta$,
\item Solve
\begin{equation}\label{eq13}
\Delta_{\rm v} f\left(\frac{r_{\rm v}}{c(r_{\rm v})r}\right)-\Delta f\left(\frac{1}{c(r_{\rm v})}\right) = 0
\end{equation}
for virial radius $r_{\rm v}$. Here $\Delta_{\rm v}$ is the virial overdensity that we find numerically solving the spherical tophat collapse model (fitting formulae for $\Delta_{\rm v}$ for some cosmological models are given in \citet{1998ApJ...495...80B}), 
\item Solve
\begin{equation}\label{eq14}
\Delta_{\rm v} f\left(\frac{r_{\rm v}}{c(r_{\rm v})r'}\right)-\Delta' f\left(\frac{1}{c(r_{\rm v})}\right) = 0
\end{equation}
for $r'$,
\item From Eq. (\ref{eq11}) find $M'$ corresponding to the radius $r'$ and overdensity $\Delta'$.
\end{enumerate}

\begin{figure}
\centering
\includegraphics[width=\plotwd]
{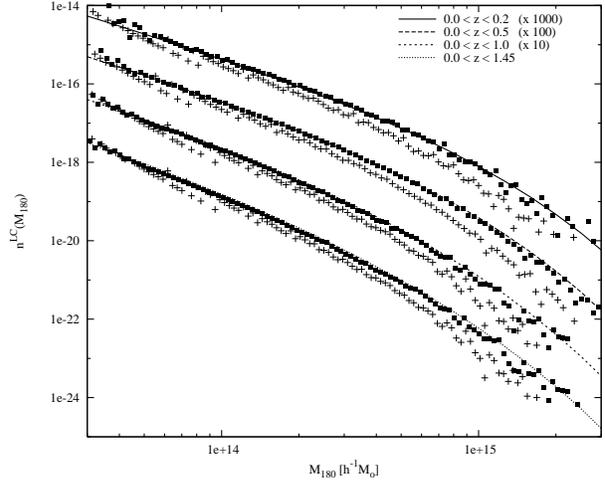}
\caption{Light-cone mass functions for different redshift intervals. For clarity the curves have been shifted by the factors given in the legend. Boxes show simulation results with the applied mass conversion from $M^{\rm{C}}_{200}$ to $M_{180}$ whereas crosses are the results without any conversion.}
\label{fig2}
\end{figure}

The results of this mass conversion from $M^{\rm{C}}_{200}$ to $M_{180}$ are shown in Fig. \ref{fig1} as a fractional increase in mass $f=M_{180}/M^{\rm{C}}_{200}-1$ for different redshifts. We see that especially for low redshift clusters this mass change can reach up to $50\%$. In Fig. \ref{fig2} we demonstrate the importance of the mass conversion in order to get agreement with the analytical mass function calculations. Here the mass function on a light-cone was calculated as follows:
\begin{equation}\label{eq15}
n^{\rm{LC}}(M_{180},<z)=\frac{\int \limits_{0}^{\rm{z}}n(M_{180},z)\frac{{\rm d}V_{\rm c}}{{\rm d}z}{\rm d}z}{\int \limits_{0}^{\rm{z}}\frac{{\rm d}V_{\rm c}}{{\rm d}z}{\rm d}z}.
\end{equation}
Here ${\rm d}V_{\rm c}$ is a comoving volume element and $n(M,z)$ is a mass function as described in \citet{1999MNRAS.308..119S}, which is known to give a very good description of N-body results \citep{2001MNRAS.321..372J}.

\begin{figure}
\centering
\includegraphics[width=\plotwd]
{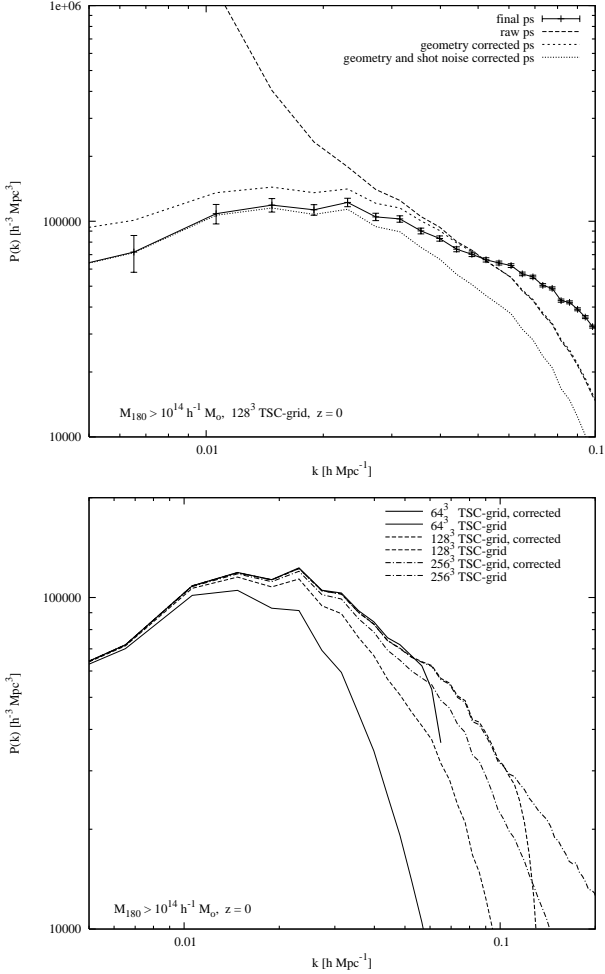}
\caption{Upper panel: various corrections applied to reach the final power spectrum estimate. Lower panel: results of the consistency test for the ``sharpening'' scheme using different grid sizes.}
\label{fig3}
\end{figure}

\begin{figure}
\centering
\includegraphics[width=\plotwd]
{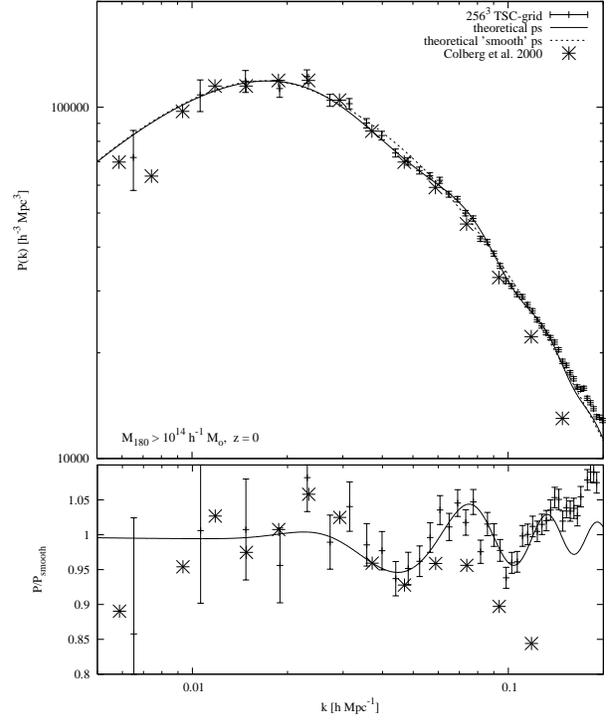}
\caption{Upper panel: Power spectrum of clusters more massive than $1.0\cdot10^{14} h^{-1}M_{\odot}$ from $z=0$ simulation box. Points with errorbars show the power spectrum of clusters inside a spherical subvolume with comoving radius of $1500$ $h^{-1}\,\mathrm{Mpc}$ extracted from the full simulation box. Stars present results obtained by \citet{2000MNRAS.319..209C} and solid/dashed lines are model power spectra with/without acoustic oscillations. Lower panel: as above, except all the curves have been divided by the ``smooth'' model without acoustic oscillations.}
\label{fig4}
\end{figure}

\begin{figure*}
\centering
\includegraphics[width=\plotwdtwo]
{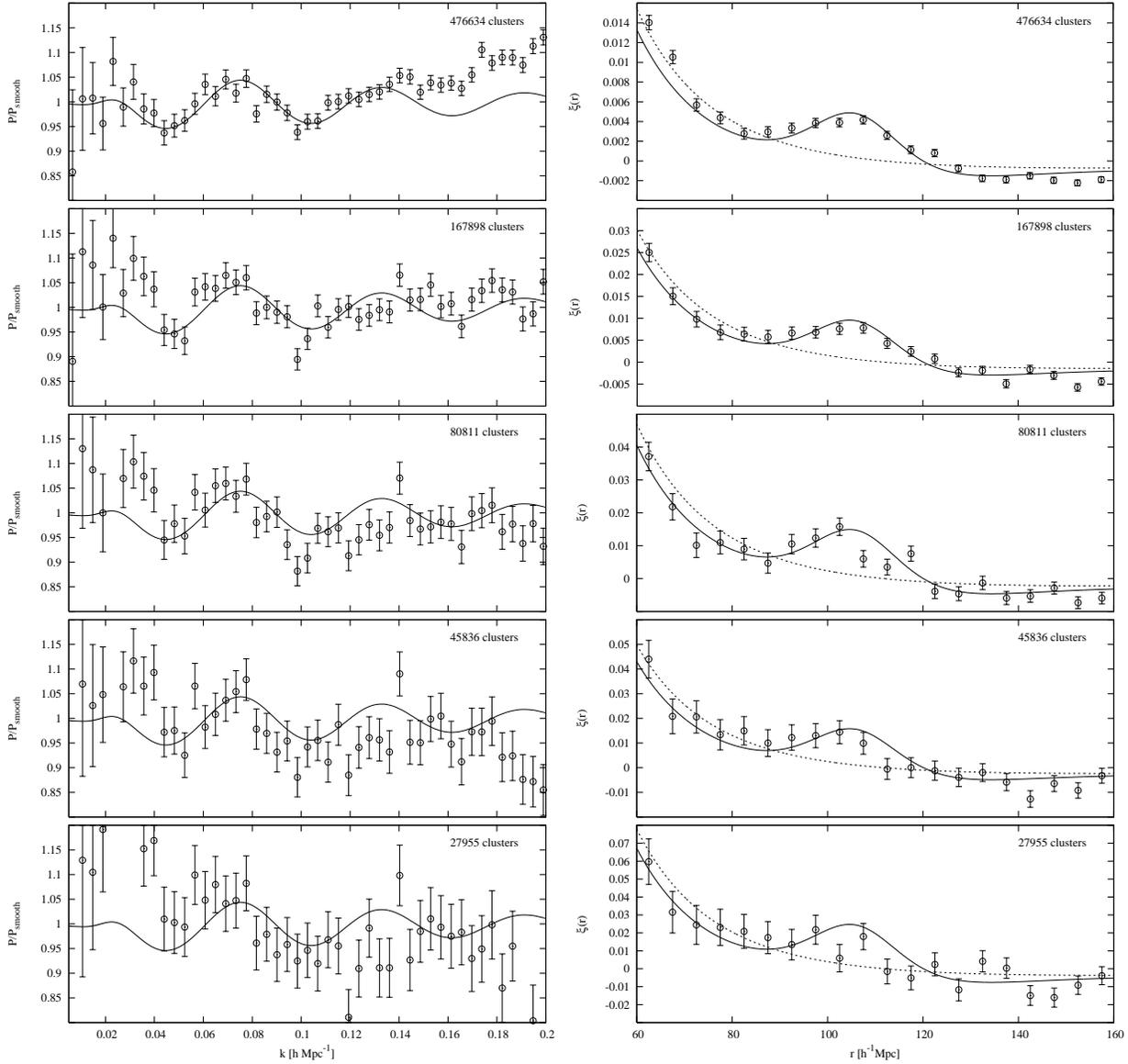}
\caption{Power spectra (left panels) and correlation functions (right panels) for the lower mass cutoffs of $1.0\cdot10^{14}$, $2.0\cdot10^{14}$, $3.0\cdot10^{14}$, $4.0\cdot10^{14}$ and $5.0\cdot10^{14}$ $h^{-1}M_{\odot}$ (friend-of-friend masses). Power spectra have been divided by the model spectra without acoustic features. The number of objects inside a spherical survey volume reaching $z=0.58$ is given in each panel. Solid/dotted lines show theoretical models with/without baryonic oscillations.}
\label{fig4b}
\end{figure*}

In the upper panel of Fig. \ref{fig3} we show various corrections needed to achieve a reliable estimate of the power spectrum of the underlying cluster distribution while the lower panel demonstrates the consistency of the applied ``sharpening'' scheme. Here we have used the $z=0$ cluster catalog to allow for a comparison with the results presented in \citet{2000MNRAS.319..209C}. The lower mass for the cluster selection was taken to be $1.0\cdot10^{14} h^{-1}M_{\odot}$ in order to get the total number of objects equal to $\sim 915,000$ as was used in \citet{2000MNRAS.319..209C}. Also we have selected a spherical volume out of the full box to test how well the geometry correction works. The results of this comparison are given in Fig. \ref{fig4}. We see that the \citet{2000MNRAS.319..209C} power spectrum agrees with our calculations at the largest scales; however, for the smaller scales it drops below our results. We suspect that their correction for the grid smoothing effect was insufficient, although in their paper they do not describe how the power spectrum was calculated. As can be seen from the figure the shape of our cluster power spectrum agrees very well with the linear theory matter power spectrum up to the scale $k \sim 0.15\,h\,\mathrm{Mpc}^{-1}$. Clearly with such a large number of clusters ($\sim 477,000$ inside our spherical volume) baryonic oscillations are easily detectable and the corresponding ``smooth'' model without them is disfavored. The theoretical matter power spectra were calculated as described in \citet{1998ApJ...496..605E}. Using the $z=0$ cluster catalog we also calculate power spectra and two-point correlation functions for various lower mass cutoffs. These results are presented in Fig. \ref{fig4b} where the left-hand panels show power spectra divided by the smooth model without baryonic oscillations and right-hand panels the respective correlation functions. Here the uppermost power spectrum is the same as the one given in Fig. \ref{fig4}. Solid/dotted lines show theoretical models with/without baryonic oscillations. Correlation functions were calculated using the estimator given by \citet{1993ApJ...412...64L}:    \begin{equation}
\xi (r) = \frac{DD - 2DR + RR}{RR},
\end{equation}      
which has minimal variance for a Poisson process. Here DD, DR and RR represent the respective normalized data-data, data-random and random-random pair counts in a given distance range. Random catalogs were generated with ten times the number of objects in the main catalogs. The survey geometry was again taken to be a spherical volume reaching redshift $z = 0.58$. The number of objects corresponding to the lower mass cutoffs of $1.0\cdot10^{14}$, $2.0\cdot10^{14}$, $3.0\cdot10^{14}$, $4.0\cdot10^{14}$ and $5.0\cdot10^{14}$ $h^{-1}M_{\odot}$ (friend-of-friend masses) were respectively $476,634$, $167,898$, $80,811$, $45,836$ and $27,955$. For the correlation function we have shown only a simple Poissonian errors:\begin{equation}
\Delta \xi \simeq \frac{1 + \xi}{\sqrt{DD}}.
\end{equation}
(Errors due to the terms DR and RR can be neglected because of the much larger number of available pairs.) These errorbars are an underestimate of the true variance, being only exact for the Poissonian point process. To estimate true errors, one needs to have a knowledge about the 3- and 4-point correlation function of the clustering pattern. We are not going to elaborate further on these issues since the following analysis is based solely on the power spectrum. From Fig. \ref{fig4b} we see that baryonic features are visible down to the case with the lowest number of clusters. The only exception is the correlation function with the lowest number of objects which shows rather noisy behavior near the expected baryonic bump. For the least massive systems, due to the nonlinear evolution, the power spectrum starts to rise at $k \gtrsim 0.15\,h\,\mathrm{Mpc}^{-1}$. For the more massive systems, on the other hand, the opposite trend is visible i.e. a decrease of power. This is due to the cluster formation which can be viewed as a smoothing filter acting on an initial density perturbation field.

\subsection{Comparison with the analytical description: accuracy of the biasing scheme}
The analytical power spectrum of clusters with masses $M>M_{\rm low}$ on a light-cone $P^{\rm{LC}}_{\rm c}(k;>M_{\rm low})$ is calculated as presented in \citet{1999ApJ...527..488Y} with a slight modification to allow for various weight functions, so:
\begin{equation}\label{eq16}
P^{\rm{LC}}_{\rm c}(k;>M_{\rm low})=\frac{\int \limits_{z_{\rm min}}^{z_{\rm max}} W^2(z)P_{\rm c}(k;>M_{\rm low},z)\frac{{\rm d}V_{\rm c}}{{\rm d}z}{\rm d}z}{\int \limits_{z_{\rm min}}^{z_{\rm max}} W^2(z)\frac{{\rm d}V_{\rm c}}{{\rm d}z}{\rm d}z},
\end{equation}
where the weight function $W(z)$ is given in Eq.(\ref{eq8}). There the number density of objects is provided by the cumulative mass function at redshift $z$:
\begin{equation}\label{eq17}
n(>M_{\rm low},z) = \int \limits_{M_{\rm low}}^{\infty}n(M,z){\rm d}M.
\end{equation}
The power spectrum of clusters more massive than $M_{\rm low}$ at redshift $z$ is given as:
\begin{equation}\label{eq18}
P_{\rm c}(k;>M_{\rm low},z)=D^2_+(z)b^2_{\rm eff}(>M_{\rm low},z)P(k,z=0),
\end{equation}
where the effective bias parameter:
\begin{equation}\label{eq19}
b_{\rm eff}(>M_{\rm low},z)=\frac{\int \limits_{M_{\rm low}}^{\infty}b(M,z)n(M,z){\rm d}M}{n(>M_{\rm low},z)}.
\end{equation}
$D_+(z)$ is the growing mode of linear density fluctuations normalized such that $D_+(z=0)=1$ and $P(k,z=0)$ is the matter power spectrum at the current epoch, which is calculated using the transfer functions presented in \citet{1998ApJ...496..605E}.
   
\begin{figure}
\centering
\includegraphics[width=\plotwd]
{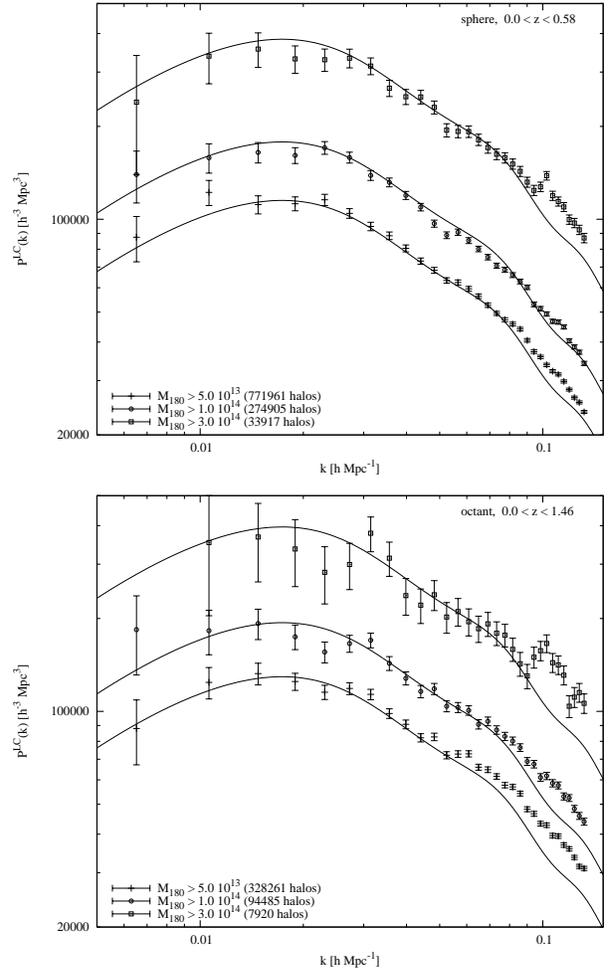}
\caption{Upper panel: light-cone power spectra for SphereB output of VIRGO $\Lambda$CDM Hubble Volume simulation for different lower mass cutoffs (see legend). Lower panel: same as above but for OctantB output.}
\label{fig5}
\end{figure}

For the mass function $n(M,z)$ and bias parameter $b(M,z)$ we use both Press-Schechter (PS) \citep{1974ApJ...187..425P} and Sheth-Tormen (ST) \citep{1999MNRAS.308..119S} prescriptions. It is well known that PS mass function underpredicts the number density of massive objects \citep{1999MNRAS.308..119S}. PS overestimates while ST underestimates the bias parameter for massive halos (especially at larger redshifts) \citep{1999MNRAS.308..119S}. PS underestimation of number density turns out to be approximately compensated for by its overestimation of bias parameter, and as such, we get the best agreement with the numerical light-cone power spectra using a plain PS approach. This is demonstrated in Fig. \ref{fig5} where we show the light-cone power spectra for various values of lower mass cutoff $M_{\rm low}$. Results in the upper panel apply for the SphereB cluster catalog (reaching redshift $z \sim 0.58$) whereas the ones on the lower part of the figure correspond to the OctantB catalog (reaching $z \sim 1.46$) of VIRGO Hubble Volume simulation outputs. We obtained the best agreement if clusters were selected using $M_{180}$, but in bias calculations virial mass $M_{\rm vir}$ was used instead.

Overall the agreement between the numerical results and an analytical description is better than $20\%$.

\begin{figure}
\centering
\includegraphics[width=\plotwd]
{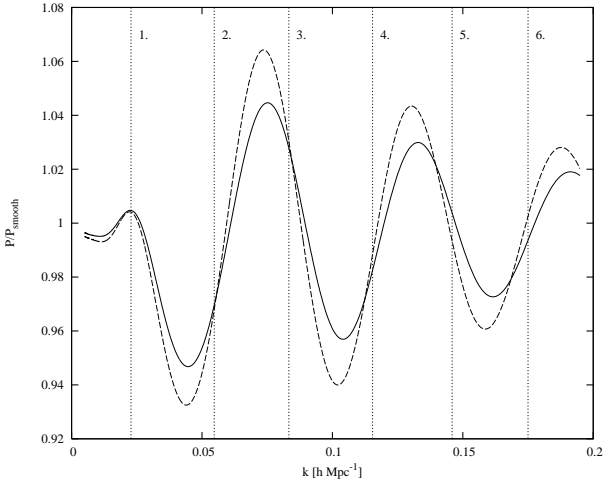}
\caption{Theoretical dark matter power spectra with a smooth component divided out (using transfer functions as given by \citet{1998ApJ...496..605E}) for the WMAP ``concordance'' model (solid line) and for the model used in VIRGO Hubble Volume simulations (dashed line). The numbered vertical lines show the locations of the corresponding peaks in the CMB angular power spectrum e.g. 1. corresponds to the 1st acoustic peak etc.}
\label{fig12}
\end{figure}

\section{SZ clusters and baryonic oscillations}
The ``concordance'' cosmological model predicts oscillations in the matter power spectrum with a relative amplitude of $\sim 5 \%$ (see Fig. \ref{fig12}).
 The correspondance of the peaks in the matter power spectrum to the ones in the CMB angular power spectrum is also given in Fig. \ref{fig12} (The first vertical line represents the position of the 1st CMB acoustic peak etc.). We see that on small scales the corresponding features are out of phase. This is due to the so-called velocity overshoot, meaning that at those scales the growing mode of density fluctuations is mostly sourced by the velocity perturbations \citep{1970Ap&SS...7....3S,1998ApJ...496..605E}. At larger scales, on the other hand, fluctuations in the density provide a dominant source term and so the corresponding features in the power spectra are in phase. Each rise and fall in the matter power spectrum corresponds to twice as many features in the CMB angular spectrum. This is a generic property of the models with a dominating CDM component. Purely baryonic models on the other hand would have oscillations with the same frequency as in the angular spectrum of the CMB. \footnote{http://cmb.as.arizona.edu/$\sim$eisenste/acousticpeak/}
In order to see these features in the matter power spectrum one needs a ``tracer'' population of objects whose clustering properties with respect to the underlying dark matter distribution are reasonably well understood. These objects should have high enough number density to reduce discreteness noise on one hand, but on the other hand they should fill as large of a comoving volume as possible to decrease cosmic variance. In general one wants to find an optimal solution of these two degrees of freedom \footnote{One also has to consider the steepness of the luminosity function of those objects.}, as to maximize the obtainable effective volume (see Eq. (\ref{eq7})) for a fixed observational effort. Currently the largest effective volume amongst all of the available surveys is provided by the SDSS LRG sample \citep{astro-ph/0501171}.
The analysis of this sample yielded a detection of a clear acoustic feature in the spatial two point correlation function. Future projects such as the K.A.O.S.\footnote{http://www.noao.edu/kaos/} galaxy redshift survey has as one of its main scientific targets the detection of baryon oscillations in the spatial clustering of high$-z$ galaxies. The possibility of using the aforementioned galaxy redshift surveys to measure the sound horizon has been discussed in several papers e.g. \citet{2003ApJ...594..665B,2003PhRvD..68h3504L,2003ApJ...598..720S,2003PhRvD..68f3004H} and a similar discussion in the context of photometric redshift surveys is given in \citet{astro-ph/0411713} (see also the discussion in \citealt{2003ApJ...598..720S}).

Here instead of galaxies we discuss the possibility of using SZ-selected galaxy clusters for that purpose. Some calculations related to the SPT-type of SZ survey were also presented in \citet{2003PhRvD..68f3004H}. It is clear from Figs. \ref{fig4} and \ref{fig12} that with wide field galaxy cluster surveys we should be especially sensitive to the scales that correspond to the 2nd and 3rd acoustic peaks in the CMB angular power spectrum. 

A few advantages of using galaxy clusters compared to the galaxies are:
\begin{itemize}
\item With relatively small cluster samples it is possible to probe large cosmological volumes (thus reducing cosmic variance).
\item The clustering signal of galaxy clusters is amplified with respect that of galaxies.
\item The relation with respect to the underlying dark matter field is rather well understood and also redshift space distortions are managable since ``fingers of god'' could be avoided.\footnote{This is only true when we have enough galaxy redshifts per cluster, so that one can average down to something close to the center of mass velocity. One might also try to exploit the fact that the bright central cD galaxies have small velocities with respect to the rest of the cluster (see the simulation results by \citet{2003ApJ...593....1B}). Even if one is able to find a good estimate for the center of mass velocity for each cluster in the sample, the redshift space distortions on reasonably large scales would still deviate from the simple linear prediction \citep{2004PhRvD..70h3007S}.}
\end{itemize}
The biggest disadvantage is a rather low number density i.e. high shot noise contribution.
\subsection{SZ-selected clusters. Mass-observable relations}
\begin{figure}
\centering
\includegraphics[width=\plotwd]
{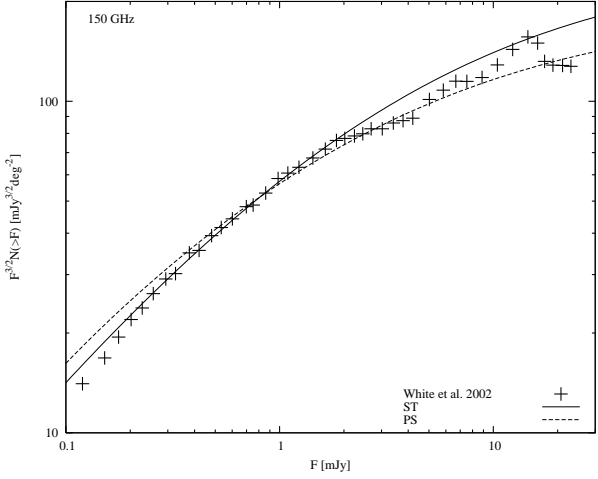}
\caption{Cumulative number counts of SZ clusters for observing frequency $150$ GHz. Crosses show simulation results by \citet{2002ApJ...579...16W} whereas solid/dashed lines correspond to analytical results assuming a ST/PS mass function.}
\label{fig6}
\end{figure}

In order to compare observations with the models one has to establish mass-observable relations and also specify survey selection criteria. Here for the sake of simplicity we assume that all the clusters remain unresolved i.e. we assume that our sample is effectively flux-selected. This is a rather good approximation for the case of PLANCK, but for surveys like ACT and SPT extra complications will arise since part of the cluster population will be resolved and so the selection function has one additional degree of freedom, namely surface brightness. As our aim here is not to give any detailed predictions for a particular survey these assumptions seem to be quite reasonable.
The change in detected flux towards a galaxy cluster due to the thermal SZ effect can be expressed as:
\begin{equation}\label{eq20}
F(M,z,x)=\frac{I_0\sigma_{\rm T}}{\mu_{\rm e}m_{\rm p}m_{\rm e}c^2}\cdot\frac{g(x)f_{\rm b}M\cdot kT(M,z)}{d_{\rm A}(z)^2},
\end{equation}
where $I_0=2(kT_{\rm cmb})^3/(hc)^2 \simeq 2.7\cdot10^{11} \mathrm{mJy/sr}$, $f_{\rm b}$ is the cluster baryonic fraction which we take to be equal to the cosmic average $\Omega_{\rm b}/\Omega_{\rm m}$, $d_{\rm A}(z)$ is angular diameter distance to the cluster, $\mu_{\rm e}=2/(1+X)$ for the case of fully ionized plasma with negligible metallicity (we take $\mu_{\rm e}=1.14$). The spectral function $g(x)$ is given as follows \citep{1980ARA&A..18..537S}:
\begin{equation}\label{eq21}
g(x)=\left[x\coth(\frac{x}{2})-4\right]\cdot\frac{x^4\rm{e}^x}{(\rm{e}^x-1)^2},
\end{equation}
where the dimensionless frequency $x=h\nu/kT_{\rm cmb}\simeq0.0176\cdot\nu\mathrm{(GHz)}$. For the mass-temperature relation we assume a simple virial scaling \citep{1998ApJ...495...80B}:
\begin{equation}\label{eq22}
kT(M,z)=A\cdot\left[\Delta_{\rm c}(z)E(z)^2\right]^{\frac{1}{3}}M^{\frac{2}{3}}.
\end{equation}
Here $\Delta_{\rm c}(z)$ is a critical collapse overdensity with respect to the critical density at redshift $z$ and $E(z)=H(z)/H_0$. The normalizing constant $A$ is determined so as to obtain a good match for the SZ cluster number counts from the state-of-the-art hydrodynamical simulations by \citet{2002ApJ...579...16W}. These simulations included gas cooling processes and also feedback from supernovae and galactic winds. If we measure $kT$ in keV and $M$ in units of $h^{-1}M_\odot$, then a good fit can be obtained if $A\simeq1.0\cdot10^{-10}$ as seen in Fig. \ref{fig6}. Here we present results both for ST and PS mass functions.

In reality the mass-observable relations are currently rather poorly known but one may argue that planned surveys with the yields of tens of thousands of galaxy clusters have significant power for ``self-calibration'' \citep{2004ApJ...613...41M}. As also shown in \citet{2004ApJ...613...41M} a much better approach would be to establish these scaling laws using external mass determinations (e.g. through lensing studies) for a subset of a complete sample. For a more precise modeling of the selection effects one also has to consider scatter around these mean relations. These issues can be settled once we obtain a real sample. Moreover, the clustering as compared to the number count of objects is much less sensitive to the uncertainties in the precise knowledge of the selection effects. Here the selection effects enter when relating the clustering of tracer objects to the underlying dark matter i.e. while determining the effective bias of objects. As it turns out (see Sec. 4), future large cluster samples are able to provide a good estimate of the effective bias themselves through the redshift space distortions.    

\subsection{Accuracy of the power spectrum determination}

\begin{figure}
\centering
\includegraphics[width=\plotwd]
{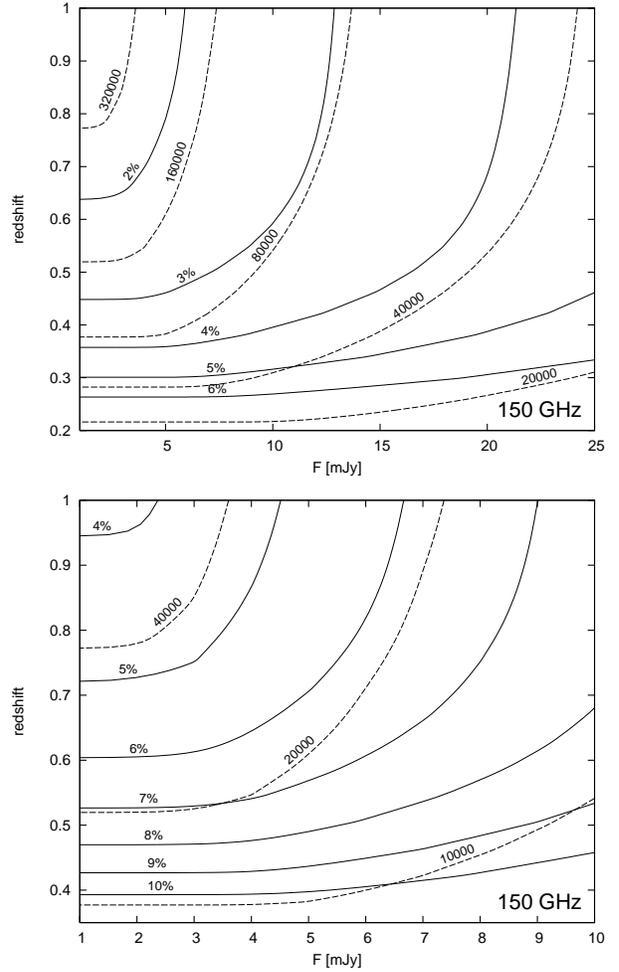}
\caption{Upper panel: $\Delta P/P$ (given in \%) and number of clusters for flux-limited surveys with various sensitivities and survey depths at $150$ GHz assuming full sky coverage. Solid lines show $\Delta P/P$ isocontours whereas constant number contours are given with dashed lines. Lower panel: the same as above only for one octant of the sky.}
\label{fig7}
\end{figure}

We investigate how well we can determine the power spectrum with SZ surveys having various sensitivity limits, specifically concentrating on the range $1\ldots25$ mJy. Our results for $\Delta P /P$ (or equivalently for the effective volume, see Eq. (\ref{eq6})) and for the number of detectable clusters are given in Fig. \ref{fig7}. Here the upper panel assumes full sky coverage and the lower one applies for one octant of the sky. With the solid lines we have plotted the fractional accuracy $\Delta P/P$ achieved for different lower flux limits and follow-up survey depths applying the FKP weighting scheme. The wavenumber $k$ in the calculations was taken to be $0.05\,h\,\mathrm{Mpc}^{-1}$, which is close to the first major acoustic feature in the expected matter power spectrum (see Fig. \ref{fig12}). The bin width $\Delta k=0.005\,h\,\mathrm{Mpc}^{-1}$ is large enough so that even for the shallowest surveys reaching only $z\sim0.2$ the power spectrum bins can be assumed to be independent. Here and also in the following we perform analytical calculations for the observational frequency $150$ GHz and assume cosmological parameters consistent with the WMAP ``concordance'' model \citep{2003ApJS..148..175S}.

Light-cone power spectra for SZ clusters are calculated using Eq. (\ref{eq16}) where the lower integration boundary $M_{\rm{low}}(F,z)$ is given by Eq. (\ref{eq20}) and (\ref{eq22}). Then $\Delta P/P$ is found using Eq. (\ref{eq6}),(\ref{eq7}) and (\ref{eq8}) and the mean underlying number density $\bar{n}(z)$ is given by Eq. (\ref{eq17}). Also we have taken into account the increase of the isotropized power spectrum due to linear redshift space distortions by a factor of $1+2\beta/3+\beta^2/5$ \citep{1987MNRAS.227....1K}, where  $\beta=\frac{1}{b_{\rm eff}}\cdot\frac{{\rm d}\ln D_+}{{\rm d}\ln a}$.

The solid lines in Fig. \ref{fig7} starting from below correspond to $\Delta P/P$ values of $6\%$, $5\%$, $4\%, \ldots$ in the upper panel and $10\%$, $9\%$, $8\%, \ldots$ in the lower one. With the dashed lines we have plotted the number of clusters. Moving from the lower right to the upper left each line represents a factor of two increase in number with the starting values being $20,000$ and $10,000$ in upper and lower panels, respectively. The flattening out of $\Delta P/P$ and cluster number curves at low fluxes is due to the imposed lower mass cutoff $1.0\cdot10^{14}h^{-1}M_{\odot}$. Thus below some flux limit we see all the clusters inside a specified volume that have masses above that cutoff value. 

In general, by increasing the volume of the survey we also boost the shot noise contribution due to the decreasing number density of distant objects. As seen from the figure- in the case of FKP weighting- the accuracy of power spectrum estimate does not degrade as we move to further distances since we downweight the contribution of the far away objects in such a way as to compensate for the increase in shot noise. For too small survey volumes, on the other hand, limits on achievable accuracy are set by the growing importance of the cosmic variance. The FKP weighting scheme is not strictly optimal in our case since it was derived assuming a fixed i.e. non-evolving underlying power spectrum. Certainly, for a better scheme one should weight down the contribution of the far away objects slightly more mildly since the clustering strength of these objects is higher. This type of weighting method, which is able to handle at least the case with an evolving amplitude, is presented in \citet{2004MNRAS.347..645P}. Nevertheless in the following calculations for simplicity we still apply the FKP weight function.  

\begin{figure}
\centering
\includegraphics[width=\plotwd]
{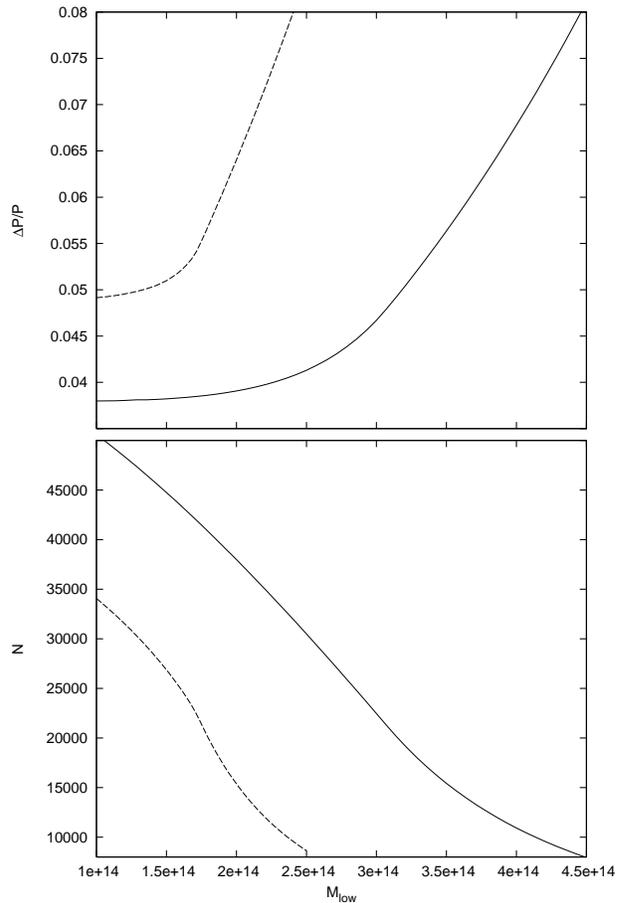}
\caption{$\Delta P/P$ (upper panel) and number of clusters (lower panel) as a function of the lower mass cutoff $M_{\rm low}$. Solid lines correspond to the full sky shallow SZ survey and dashed lines to the deep survey covering one octant of the sky.}
\label{fig9}
\end{figure}

The results in Fig. \ref{fig7} assumed that we have a full follow-up such that we are able to obtain all the redshifts of the clusters detected by an imaging survey. Also the sample was assumed to be purely flux-selected which is rather unrealistic for real experiments e.g. for PLANCK many clusters remain undetected due to rather poor angular resolution or oppositely in the case of SPT some fraction of clusters will be ``resolved out'' and a significant amount of signal will be lost. Many objects might remain undetected for these reasons. 
If the systems that are left out are low mass clusters (as in the case of PLANCK) then our power spectrum estimate might actually 
be almost as good as before since low mass systems are relatively weakly clustered and as such they do not contribute significantly to the total signal. This can be seen in Fig. \ref{fig9} where in the upper panel we have shown the influence of changing the lower mass cutoff $M_{\rm low}$ on $\Delta P/P$. On the lower panel the respective number of clusters is given. These calculations are done for two different survey types: (1) solid lines represent results for a shallow SZ survey covering the full sky and reaching redshift $z_{\rm lim}=0.6$ with a sensitivity limit $F_{\rm low}=17$ mJy at $150$ GHz, (2) dashed lines represent a deep survey (with no upper redshift cutoff imposed) covering $1/8$ of the sky with a flux limit $F_{\rm low}=5$ mJy at $150$ GHz. The first might be applicable for the case of the PLANCK mission\footnote{Using the spectral dependence of thermal SZ effect (see Eq. (\ref{eq21})) we see that the $30$ mJy sensitivity of PLANCK's $353$ GHz channel corresponds to $17$ mJy at $153$ GHz.} and the second for the SPT cluster survey. In practice the measurements will be performed in many frequency channels which helps to separate clusters from other foreground sources due to the specific frequency behavior of the thermal SZ effect. Here for simplicity we have chosen the sensitivity limits corresponding to the ``weakest'' of the channels available for SZ purposes. We see that for a full sky with the $20,000$ most massive clusters up to redshift $z\sim 0.6$, one could obtain an estimate of the power spectrum at $k=0.05\,h\,\mathrm{Mpc}^{-1}$ with a fractional error below $5\%$ while for the one octant of the sky with less than $\sim 25,000$ clusters we always stay above $5\%$ accuracy.
   
The previously described approach where we use all the data to obtain a single combined estimate of the power spectrum is appropriate if we only intend to place constraints on $\Omega_{\rm DE}$. Combining this power spectrum estimate which is sensitive to $\Omega_{\rm m}h$ with the CMB constraint on $\Omega_{\rm m}h^2$ gives us $\Omega_{\rm m}$ and $h$ separately. Additionally, knowing the geometry of the Universe from CMB measurements gives us immediately an estimate of $\Omega_{\rm DE}$. Since in the majority of the DE models DE starts to dominate relatively recently, driving the Universe furthest from the plain Einstein-de Sitter behavior, the best redshift to complement the CMB data is at $z=0$.
 
\begin{figure}
\centering
\includegraphics[width=0.55\textwidth]
{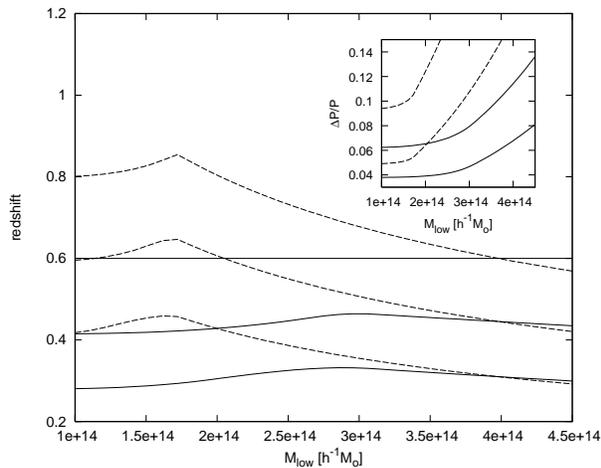}
\caption{``Equal accuracy'' redshift binning for different lower mass cutoffs. Solid (dashed) lines correspond to the shallow (deep) SZ survey with 3 (4) redshift bins. In the inset the upper curves show the relative accuracy achievable in each redshift bin while the lower lines correspond to the full sample without any binning.}
\label{fig11}
\end{figure}

On the other hand if our aim is to constrain the equation of state parameter $w$ and its possible change in time it is essential to measure the power spectrum at different redshifts. This leads to the question of how to bin up the sample in redshift? Certainly there are optimal ways of combining data, but unfortunately they all depend on the way we choose to parametrize our model for DE. Recently  \citet{2003PhRvL..90c1301H} argued that in the absence of a theoretically well motivated parametrization one should use a stepwise function with the value $w_{\rm i}$ in the $i$-th redshift bin and let the data itself determine which combinations of $w_{\rm i}$ will be well constrained. One can then reconstruct the behavior of $w$ with the redshift as a linear combination of the ``cleanest'' eigenmodes. Here we are not trying to implement that kind of general parametrization since as a first step it should be sufficient to determine an effective constant $w$ and see whether it deviates from the currently most well motivated $w=-1$. Therefore, in the following we mostly investigate the case with a constant equation of state parameter $w_0$. The redshift binning is chosen so as to get equal relative accuracies of the power spectrum in each bin. 
The results of this binning procedure for the above described two types of survey are given in Fig. \ref{fig11} with the solid lines corresponding to the shallow one. For the shallow (deep) survey we have assumed 3 (4) redshift bins. In the inset the upper curves show the relative accuracy achievable in each redshift bin while the lower lines correspond to the full sample without any binning and so coincide with the lines shown in the upper panel of Fig. \ref{fig9}. This kind of redshift division is also used in the following parameter estimation section.  

\subsection{Prospects of detecting baryonic ``wiggles''. Comparison with SDSS LRG}
The relative amplitude of acoustic oscillations in the mater power spectrum for the WMAP ``concordance'' model as well as in the power spectrum used in VIRGO Hubble Volume simulations is shown in Fig. \ref{fig12}. Here the smooth component was divided out using the fitting formulae provided by \citet{1998ApJ...496..605E}. In order to be able to detect these features the accuracy of the power spectrum determination should be of comparable size i.e. $\Delta P/P\sim5\%$. This implies that for a full sky survey one needs on the order of $25,000$ galaxy clusters inside the volume with limiting redshift $z\sim0.6$ as seen from the upper panel of Fig. \ref{fig9}. It is clear that with clusters one might hope to detect only the first few acoustic signatures, e.g. the major features at $k\sim 0.045$ and $\sim 0.075\,h\,\mathrm{Mpc}^{-1}$, since they are too rare objects to enable the sampling of the smaller scale density field.

\begin{figure}
\centering
\includegraphics[width=\plotwd]
{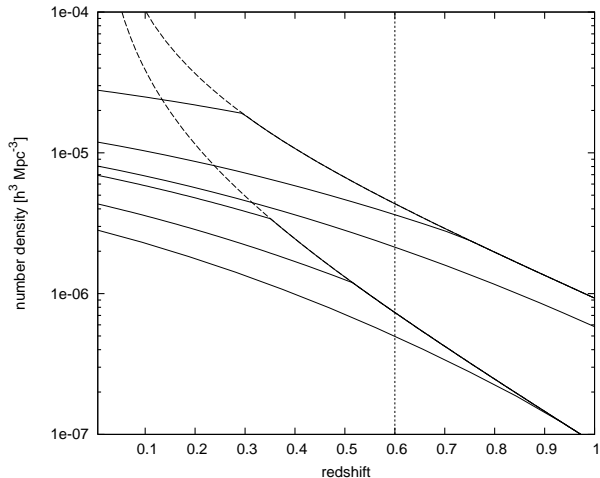}
\caption{Comoving number density of clusters as a function of redshift for the PLANCK- (lower set of curves) and SPT-like (upper set of curves) SZ surveys. Solid lines in each set correspond to the cases with $15,000$,$25,000$ and $35,000$ clusters. The vertical dotted line shows the applied upper limiting redshift for the PLANCK. Dashed lines represent purely flux-limited cases i.e. without any lower mass cuttoff imposed.}
\label{fig13}
\end{figure}

\begin{figure}
\centering
\includegraphics[width=\plotwd]
{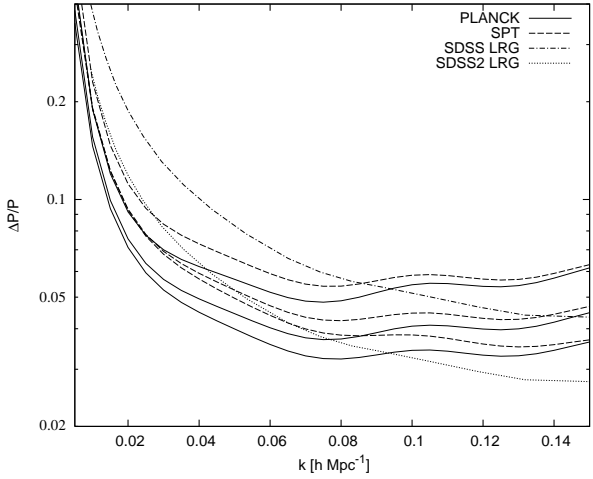}
\caption{Comparison of the ``strength'' of the future SZ cluster surveys with respect to the SDSS LRG. Lines shown for the PLANCK- and SPT-like surveys correspond to the detected cluster numbers (starting from above): $15,000$, $25,000$ and $35,000$.}
\label{fig14}
\end{figure}

Since the study of the SDSS LRG sample has led to the detection of acoustic oscillations in the spatial distribution of galaxies \citep{astro-ph/0501171}, \footnote{The detection has also been claimed using the 2dF redshift survey \citep{astro-ph/0501174}.} it would be instructive to compare the ``strength'' of this survey to the planned blank sky SZ cluster surveys like PLANCK and SPT. 
In Fig. \ref{fig13} we show the number density of clusters as a function of redshift for the above described two types of SZ survey. The upper group of lines corresponds to the SPT-like deep survey while the lower curves are for a wide and shallow survey like PLANCK. For each of the surveys we have varied the lower mass cutoff $M_{\rm low}$ so as to obtain in total $15,000$, $25,000$ and $35,000$ clusters. These three cases are shown with solid lines. Dashed lines display the pure flux-limited surveys without any lower cutoff in the mass imposed. Using these results and also taking into account the proper biasing factors as given by the square root of Eq. (\ref{eq16}) divided by $P(k,z=0)$ \footnote{The light-cone bias parameters calculated this way are $5.0$,$4.7$,$4.3$ for the PLANCK and $4.4$,$4.1$,$3.9$ for the SPT with the number of clusters $15,000$,$25,000$ and $35,0000$, respectively.} we can readily obtain $\Delta P/P$ as given by Eq. (\ref{eq6}) and (\ref{eq7}). The results of this calculation are given in Fig. \ref{fig14}. Here the solid lines correspond to the PLANCK-like and dashed ones to the SPT-like surveys. Each set of lines corresponds to the detected cluster numbers (starting from above): $15,000$, $25,000$ and $35,000$. The dash-dotted curve, showing the results for the SDSS LRG sample, is found using Eq. (\ref{eq6}) and the data for the effective volume given in Fig. 1 of \citet{astro-ph/0501171} (again $\Delta k=0.005\,h\,\mathrm{Mpc}^{-1}$ was assumed). We can see that on large scales future SZ surveys have enough strength to improve the results obtained using the SDSS LRG sample. However, the SDSS LRG sample is going to double in size within a few years as the survey is completed. The achievable $\Delta P/P$ for this final sample is shown in Fig. \ref{fig14} as a dotted line. Moreover, it seems that acoustic oscillations are able to survive at the quasilinear scales ($k \sim 0.1\ldots 0.3\,h\,\mathrm{Mpc}^{-1}$) (see Fig. \ref{fig4}) which significantly increases the amount of information available for the galaxy redshift surveys. This is also confirmed by the recent N-body simulations by \citet{2005Natur.435..629S} and \citet{astro-ph/0507338}. In order to fully exploit this information one needs a complete theoretical understanding of how nonlinear effects, redshift space distortions and nonlinear biasing influence these features. So far there have been only a few works studying these important issues (e.g. \citealt{1999MNRAS.304..851M,2005Natur.435..629S,astro-ph/0507338,astro-ph/0507307}) and we do not have a full theoretical description of them available yet. For this reason we have not attempted to incorporate the SDSS LRG sample into our Fisher matrix parameter estimation process. 

\subsection{Some remarks on SZ vs. optical cluster selection}
SZ cluster selection might be superior to the simple optical one since SZ brightness does not suffer from ordinary cosmological dimming. This allows one to obtain an approximately mass-limited sample of clusters that is spatially much more uniform than optically selected samples. For example, in Fig. \ref{fig13} the number density of clusters for both PLANCK (up to $z \sim 0.6$) and the SPT type of surveys drops approximately as $\propto z^{-2.5}$. In contrast, the number density of the optically selected SDSS LRG sample (which contains mostly galaxies that populate dense cluster environments) drops as $\propto z^{-5}$ beyond $z \sim 0.3$. So in general with SZ-selected clusters one is able to probe larger volumes. Also as a larger part of the sample is at higher redshifts the clustering signal is stronger due to the increase of the bias factor with increasing distance. (Of course the question of whether these far away parts of the sample have high enough number density in order to be useful for the clustering study depends on the specific parameters of the experiment.) The other weaknesses of the optical selection are projection effects and confusion with the background objects. As an example, comparison of the X-ray and optically selected cluster catalogues often yields a rather poor match (e.g. \citealt{2002ApJ...569..689D}), which probably signals that many of the ``optically constructed'' systems are actually false detections.      

For the SZ surveys the spectroscopic follow-up is a crucial issue, but several other studies (such as cluster number counts) also require determination of redshifts. As these investigations will be performed anyway, the clustering study can be seen as coming esentially ``for free''.  

\begin{figure*}
\centering
\includegraphics[width=\plotwdtwo]
{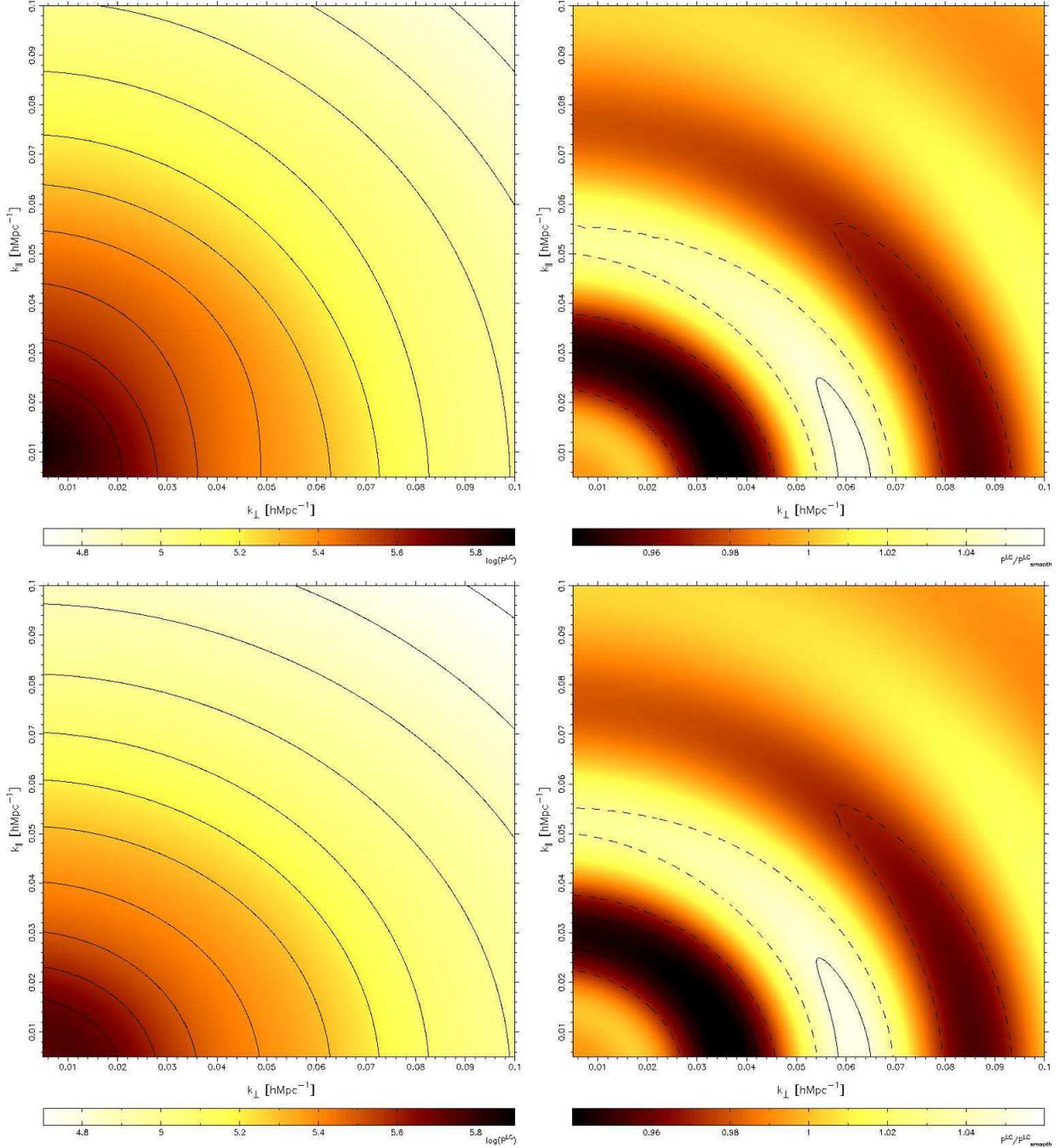}
\caption{Upper left panel: 2D light-cone power spectrum for clusters with mass above $1.0\cdot10^{14} h^{-1}M_{\odot}$ in logarithmic units. Lower left panel: as above, but the redshift distortion ``switched off''. Upper right panel: as upper left panel, only smooth component of the spectrum divided out and results shown using linear scale. Lower right panel: the same procedure applied to the lower left panel.}\label{fig15}
\end{figure*}

\begin{figure*}
\centering
\includegraphics[width=\plotwdtwo]
{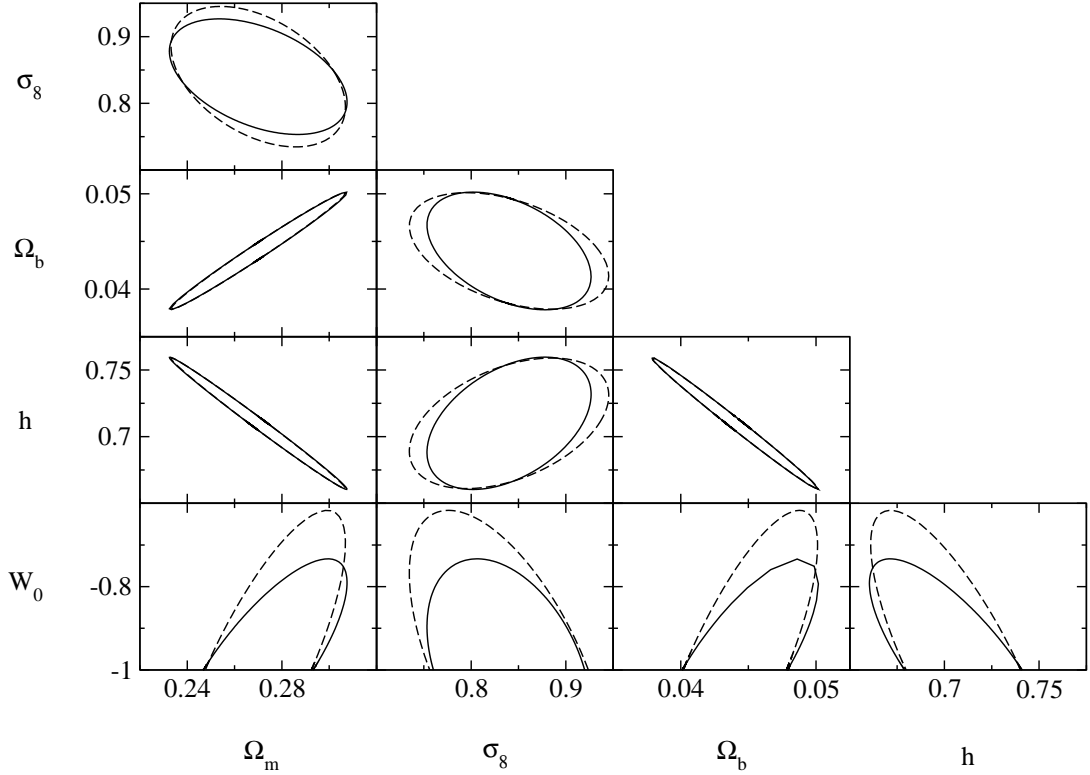}
\caption{Error ellipses for the PLANCK-like SZ survey including CMB priors as described in the text. Solid/dashed lines show results for $3/1$ redshift bin(s).}
\label{fig16}
\end{figure*}

\begin{figure*}
\centering
\includegraphics[width=\plotwdtwo]
{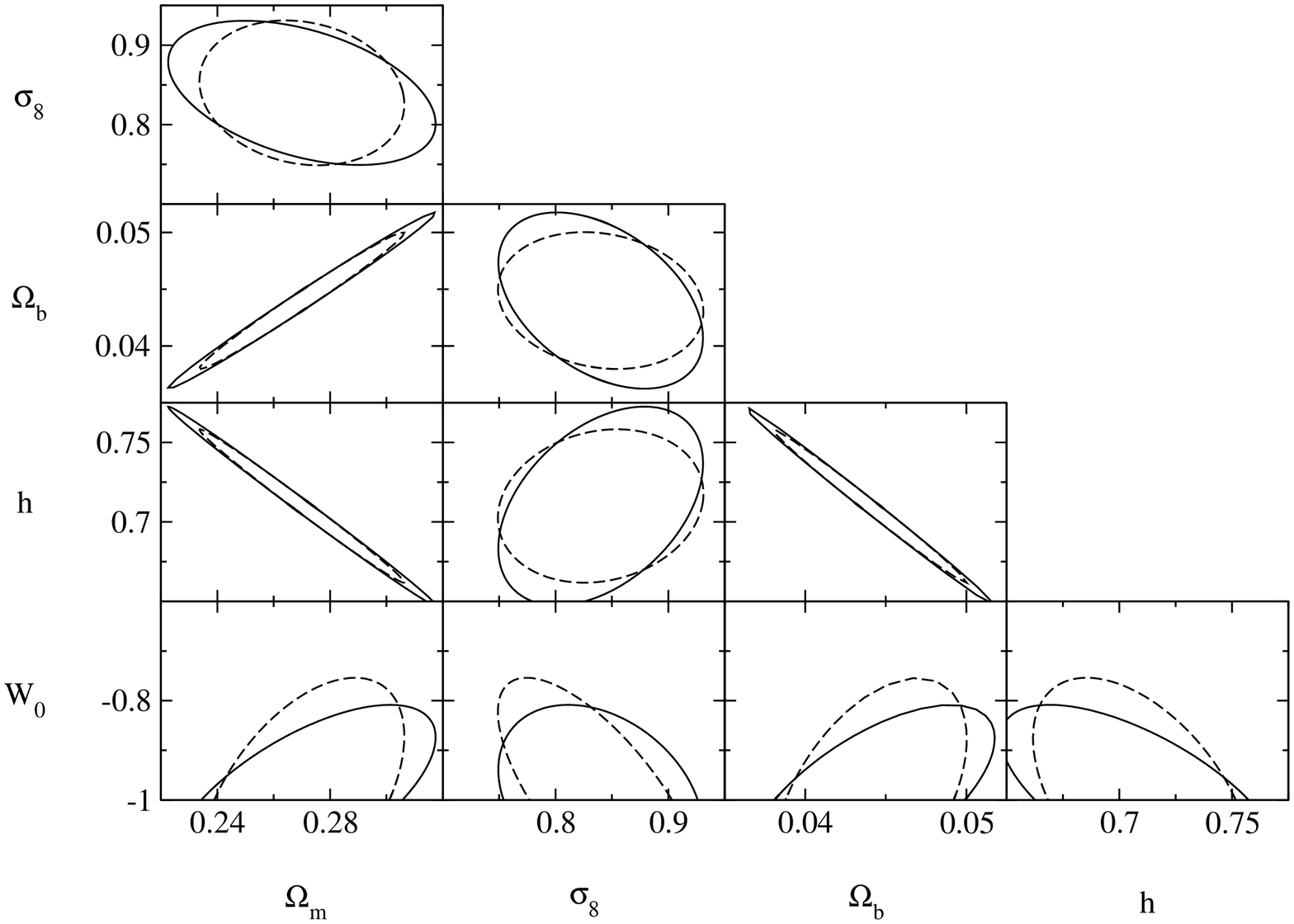}
\caption{Analog of Fig \ref{fig16} for the SPT-like survey. Solid/dashed lines show results for $4/1$ redshift bin(s).}
\label{fig17}
\end{figure*}

\section{Constraints on Dark Energy}
\subsection{2D power spectrum on a light-cone}
Having obtained a parametrized (and well calibrated) analytical model for the light-cone power spectra of SZ-selected clusters of galaxies, we can estimate the accuracy with which it is possible to recover cosmological parameters. Since the observations are done in cosmological (as opposed to the comoving) redshift space there are two additional effects one has to take into account:
\begin{enumerate}
\item The increase of power along the line of sight due to the large scale coherent inflows towards massive accretion centers. This effect is accounted for using results from linear theory.
\item Cosmological distortion due to the fact that one can directly observe only redshifts, and in order to find the corresponding comoving distances, one has to assume some cosmological model. Choosing an incorrect model will lead to distortions along and perpendicular to the line of sight \footnote{The Alcock-Paczynski test \citep{1979Natur.281..358A} is based on these cosmological distortions.}. 
\end{enumerate}
Both of these effects will, in general, lead to an anisotropic power spectrum. Thus instead of a one dimensional (isotropized) power spectrum one has to consider here a two dimensional power spectrum with components along and perpendicular to the line of sight. For simplicity, in the following we use a flat sky approximation. The 2D power spectra are calculated following the description given in \citet{2000ApJ...528...30M} with a slight modification to allow for various weight functions \footnote{In the calculations we use the FKP weight function.}. \citet{2000ApJ...528...30M} took the fiducial cosmology given by the Milne's empty universe model. Thus the comoving distance intervals along and perpendicular to the line of sight are:
\begin{equation}\label{eq23}
\Delta x_{\|}^{\rm{ref}}=\frac{c}{H_0}\Delta z,
\end{equation}
\begin{equation}\label{eq24}
\Delta x_{\bot}^{\rm{ref}}=\frac{c}{H_0}z\Delta \theta,
\end{equation}
where $\Delta z$ is redshift interval and $\Delta \theta$ angular separation between two objects. For the general FRW universe the corresponding intervals read as:
\begin{equation}\label{eq25}
\Delta x_{\|}=\frac{c}{H(z)}\Delta z,
\end{equation}
\begin{equation}\label{eq26}
\Delta x_{\bot}=d_{\rm M}(z)\Delta \theta,
\end{equation}
where $d_{\rm M}(z)=(1+z)d_{\rm A}(z)$ is the comoving transverse distance. Now, defining the shift parameters:
\begin{equation}\label{eq27}
c_{\|}(z)=\frac{\Delta x_{\|}}{\Delta x_{\|}^{\rm{ref}}}(z),
\end{equation}
\begin{equation}\label{eq28}
c_{\bot}(z)=\frac{\Delta x_{\bot}}{\Delta x_{\bot}^{\rm{ref}}}(z),
\end{equation}
we can write down the final expression for the 2D power spectrum on a light-cone:
\begin{eqnarray}\label{eq29}
\lefteqn{P^{\rm{LC,2D}}(k_{\|},k_{\bot};>M_{\rm low}) = } 
 \nonumber \\
\lefteqn{\frac{\int \limits_{z_{\rm min}}^{z_{\rm max}}{\rm d}z\frac{{\rm d}V_{\rm c}}{{\rm d}z}W^2(z)\left[1+\beta(>M_{\rm low},z)\left(\frac{k'_{\|}}{k'}\right)^2\right]^2\cdot P_{\rm c}(k';>M_{\rm low},z)}{\int \limits_{z_{\rm min}}^{z_{\rm max}}{\rm d}z\frac{{\rm d}V_{\rm c}}{{\rm d}z}W^2(z)c_{\bot}(z)^2c_{\|}(z)}}
\end{eqnarray}
where
\begin{equation}\label{eq30}
k'_{\|}=\frac{k_{\|}}{c_{\|}(z)},\quad k'_{\bot}=\frac{k_{\bot}}{c_{\bot}(z)},\quad k'=\sqrt{k'^2_{\|}+k'^2_{\bot}}
\end{equation}
and
\begin{equation}\label{eq31}
\beta(>M_{\rm low},z) = -\frac{1}{b_{\rm eff}(>M_{\rm low},z)}\cdot \frac{{\rm d}\ln D_+(z)}{{\rm d}\ln(1+z)}.
\end{equation}
The factor $c_{\bot}(z)^2c_{\|}(z)$ in the denominator of Eq. (\ref{eq29}) is the Jacobian determinant taking into account the change in a volume element. It is missing in the numerator due to the cancellation by the similar but inverse term arising from the transformation of the $k$-space volume element. The term in square brackets models the amplification due to the coherent inflows and the last term in the numerator, $P_{\rm c}(k';>M_{\rm low},z)$, is given earlier by Eq. (\ref{eq18}). In the case of SZ flux-selected clusters, $M_{\rm low}$ at each redshift for a given lower flux limit is found using Eq.(\ref{eq20}) and (\ref{eq22}).

In Fig. \ref{fig15} we present some examples of 2D power spectra calculated in the manner described above. The top left-hand panel shows the 2D light-cone power spectrum of clusters with a mass above $1.0\cdot10^{14} h^{-1}M_{\odot}$ (in logarithmic units) while the lower left-hand panel contains the same spectrum but with the redshift space distortion ``switched off''. The contours starting from the upper right corner correspond to the values $10^{4.8}$ and $10^{4.7}$ for the upper and lower panel, respectively and the step size was taken $10^{0.1}\,h^{-3}\,\mathrm{Mpc}^3$. It is clearly seen how linear redshift-space distortion boosts power along the line of sight. The cosmological distortion (in the currently selected reference model) on the other hand works in the opposite way. Since the chosen Milne model is strongly different from the $\Lambda$CDM ``concordance'' model the cosmological distortion is easily visible. We use Milne's model only for illustrative purposes but in the following parameter estimation part we change the reference model to the WMAP ``concordance'' cosmology. On the right-hand panels of the figure we have removed the smooth component of the power spectrum revealing the series of acoustic rings. Here the continuous contours correspond to $P^{\rm{LC,2D}}/P^{\rm{LC,2D}}_{\rm smooth}$ values of $0.95$ and $1.05$ while the dashed lines are for the values $0.97$ and $1.03$. It is important to note that the picture on both panels looks practically the same. This is due to the fact that cosmological and redshift-space distortions work in a different way: cosmological transformation stretches or compresses the power spectra on the plane of the figure whereas redshift distortion moves the spectra in a vertical direction. If we had smooth power spectra without any particular features then it would be extremely hard to disentangle these two types of distortions. Having the power spectra with acoustic features it is easy to isolate cosmological distortion by dividing out a smooth component. The ability to disentangle cosmological and redshift-space distortions is extremely important to extract the bias parameter from the survey in a self-consistent way. In the parameter estimation part of this section we see how much better one does with the model having acoustic oscillations as compared to the one without.

The total power spectrum measured over a broad $z$-interval is a weighted sum of differently distorted power spectra and so there will be some loss of acoustic features. The loss is stronger along the line of sight, as can be seen from Fig. \ref{fig15}. This is due to the currently chosen reference model where $\vert\frac{dc_{\|}(z)}{dz}\vert>\vert\frac{dc_{\bot}(z)}{dz}\vert$. Again, this effect is strongly pronounced because Milne's model differs strongly from the WMAP ``concordance'' cosmology for which our calculation was done.

\subsection{Parameter estimation}
In this subsection we apply a Fisher matrix forecasting techniques to study how well one can determine cosmological parameters. Since the pioneering investigations in the field of CMB anisotropies \citep{1996PhRvD..54.1332J} and galaxy redshift surveys \citep{1997PhRvL..79.3806T} these methods have gained great popularity. For a full description of the method with applications see \citet{1997ApJ...480...22T}. Following \citet{1997PhRvL..79.3806T} the Fisher matrix in the case of a 2D power spectrum can be written as (see also \citet{2003PhRvD..68f3004H}):
\begin{eqnarray}\label{eq32}
F_{\rm ij}=\sum_{\rm n}\sum_{\rm m} & &\frac{\partial \ln P^{\rm{LC,2D}}(k_{\| {\rm n}},k_{\bot {\rm m}})}{\partial \Theta_{\rm i}}\cdot 
\nonumber \\
& & \frac{V_{\rm k}^{\rm{m}} V_{\rm eff}}{2}\cdot \frac{\partial \ln P^{\rm{LC,2D}}(k_{\| {\rm n}},k_{\bot {\rm m}})}{\partial \Theta_{\rm j}}, 
\end{eqnarray}
where
\begin{equation}\label{eq33}
V_{\rm k}^{\rm{n}}=\frac{4\pi k_{\bot {\rm n}}\Delta k_{\bot}\Delta k_{\|}}{(2\pi)^3},
\end{equation}
and we take $\Delta k_{\bot} = \Delta k_{\|} = \Delta k$. $P^{\rm{LC,2D}}$ is given in Eq. (\ref{eq29}) and $V_{\rm eff}$ in Eq. (\ref{eq7}). The wavevector components $k_{\| {\rm i}}$ and $k_{\bot {\rm j}}$ form a rectangular grid with a step size $\Delta k$. We allow them to span the range $0.005 \ldots 0.1\,h\,\mathrm{Mpc}^{-1}$ and take $\Delta k = 0.005\,h\,\mathrm{Mpc}^{-1}$ i.e. we have a $20$x$20$ grid. It is not justified to go to higher wavenumbers than $0.1\,h\,\mathrm{Mpc}^{-1}$ since there already the simple linear scaling of the cluster power spectrum seems to break down (see Fig. \ref{fig5}). The parameters $\Theta_{\rm i}$ are taken to be $\Omega_{\rm m}$, $\sigma_8$, $\Omega_{\rm b}$, $h$, $w_0$ plus the bias parameters $b_{\rm n}$ in each redshift bin. Thus in the case of four redshift bins (which is the maximum number considered in our analysis) we have a total of nine free parameters. We also assume a fiducial cosmology given by the best-fit WMAP ``concordance'' model \citep{2003ApJS..148..175S} plus the dark energy in the form of the cosmological constant i.e. $w=-1$.

\begin{figure}
\centering
\includegraphics[width=\plotwd]
{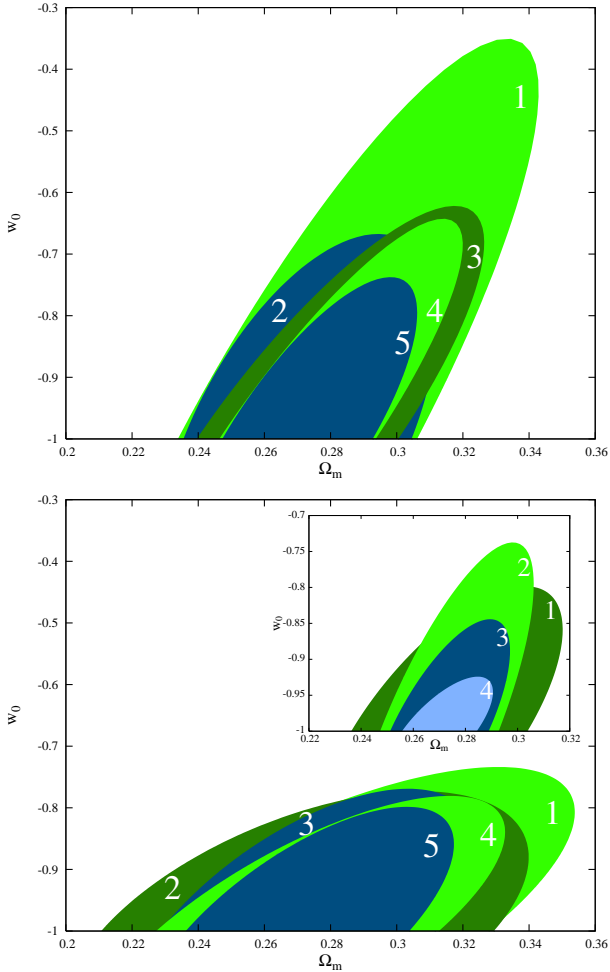}
\caption{Upper panel: Error ellipses under various assumptions for the PLANCK-like survey. Here we have shown the following cases: 1. Clustering signal of the model without baryonic oscillations + CMB priors, 2. \#1 + prior on bias, 3. Clustering signal of the ``wiggly'' model, 4. \#3 + CMB priors, 5. \#4 + prior on bias. Lower panel: as above, except for the SPT-like survey and with a changed order of the cases \#2 and \#3. Lower panel inset: constraints obtainable with: 1. SPT, 2. PLANCK, 3. SPT + PLANCK, 4. The survey with SPT characteristics covering the full sky. All of the cases here have CMB and bias priors added.}
\label{fig18}
\end{figure}

Here we perform calculations for two types of SZ survey. The first one that should serve as a prototype for PLANCK is a shallow survey that covers the full sky with a sensitivity limit of $F_{\rm low}=30$ mJy at $353$ GHz \footnote{http://astro.estec.esa.nl/Planck}. We also apply a lower mass cutoff of $M_{\rm low}=2.75\cdot10^{14} h^{-1}M_{\odot}$ resulting in total $\sim 25,000$ clusters up to the applied limiting redshift $z \simeq 0.6$ (see Fig. \ref{fig9} lower panel). The calculations are done for one redshift bin spanning $z=0 \ldots0.58$ and also for three bins: $z_1=0 \ldots 0.33$, $z_2=0.33 \ldots 0.45$, $z_3=0.45 \ldots 0.58$. The achieved $\Delta P/P$ values at $k=0.05\,h\,\mathrm{Mpc}^{-1}$ are $4.3\%$ for the one bin case and $7.5\%$ for each of the bins in the three bin case. The other deep and rather ``narrow'' survey should mimic the performance of SPT. Here we have taken $F_{\rm low}=5$ mJy at $150$ GHz and $M_{\rm low}=1.5\cdot10^{14} h^{-1}M_{\odot}$ (see \citet{2004SPIE.5498...11R,2004ApJ...613...41M}). Moreover, we assumed that the survey is capable of covering one octant of the sky and here we have not applied any high redshift cutoff. As for the PLANCK-like survey two different cases are considered here. In the first case while combining all the data to one common power spectrum we are able to achieve $\Delta P/P \simeq 5.1\%$, whereas in the second case with the four redshift bins: $z_1=0 \ldots 0.45$, $z_2=0.45 \ldots 0.63$, $z_3=0.63 \ldots 0.83$, $z_4=0.83 \ldots$ we find a $\Delta P/P \simeq 9.9\%$ in each redshift bin. The total number of objects detected with this type of survey would be $\sim 27,000$.

\begin{table*}
\caption{Principal components as given in Eq. (\ref{eq34}) for the ``clustering only'' case for PLANCK-like, SPT-like and for the combined survey.}
\label{tab1}
\begin{tabular}{l||c|l|rrrrr}
& i & $\lambda^{\rm{i}}$ & $e^{\rm{i}}_1$ & $e^{\rm{i}}_2$ & $e^{\rm{i}}_3$ & $e^{\rm{i}}_4$ & $e^{\rm{i}}_5$ \\
\hline
\hline
& $1$ & $1.6\cdot10^{3}$ & $0.74$ & $-0.0064$ & $-0.24$ & $0.62$ & $0.090$\\
PLANCK & $2$ & $1.7\cdot10^{2}$ & $-0.48$ & $-0.015$ & $-0.43$ & $0.49$ & $-0.58$\\
($3$ $z$-bins) & $3$ & $32.$ & $0.38$ & $-0.53$ & $-0.23$ & $-0.46$ & $-0.56$\\
& $4$ & $18.$ & $0.027$ & $0.41$ & $-0.80$ & $-0.37$ & $0.26$\\
& $5$ & $5.9$ & $0.27$ & $0.74$ & $0.28$ & $-0.13$ & $-0.53$\\
\hline
\hline
& $1$ & $9.5\cdot10^{2}$ & $0.71$ & $-0.094$ & $-0.23$ & $0.63$ & $0.21$\\
SPT & $2$ & $85.$ & $0.15$ & $-0.088$ & $0.29$ & $-0.36$ & $0.87$\\
($4$ $z$-bins)& $3$ & $54.$ & $-0.012$ & $-0.99$ & $-0.032$ & $-0.10$ & $-0.13$\\
& $4$ & $17.$ & $0.69$ & $0.094$ & $0.24$ & $-0.54$ & $-0.41$\\
& $5$ & $4.6$ & $0.049$ & $0.057$ & $-0.90$ & $-0.42$ & $0.12$\\
\hline
\hline
& $1$ & $2.6\cdot10^{3}$ & $0.73$ & $-0.038$ & $-0.24$ & $0.63$ & $0.13$\\
PLANCK+ & $2$ & $2.5\cdot10^{2}$ & $-0.38$ & $0.044$ & $-0.39$ & $0.45$ & $-0.70$\\
SPT & $3$ & $78.$ & $0.18$ & $-0.92$ & $-0.027$ & $-0.22$ & $-0.28$\\
& $4$ & $52.$ & $0.54$ & $0.39$ & $0.087$ & $-0.44$ & $-0.60$\\
& $5$ & $24.$ & $0.021$ & $0.057$ & $-0.89$ & $-0.40$ & $0.22$\\
\end{tabular}
\end{table*}

The results in Fig. \ref{fig16} and \ref{fig17} present constraints for the five cosmological parameters: $\Omega_{\rm m}$, $\sigma_8$, $\Omega_{\rm b}$, $h$ and $w_0$. Here we have added priors to $\Omega_{\rm m}h^2$ and $\Omega_{\rm b}h^2$ from the CMB angular power spectrum studies, and also a prior to the bias parameters. In order to be able to compare our results to the ones given in \citet{2003PhRvD..68f3004H} the fractional errors of $0.01$ for $\Omega_{\rm m}h^2$ and $\Omega_{\rm b}h^2$ (this should be achievable with the PLANCK mission \citep{2002PhRvD..65b3003H}) were similarly assumed. These constraints are easily ``rotated'' to the frame used here since under coordinate transformations the Fisher matrix transforms as a second rank tensor. Moreover, we have restricted our calculations to the flat models only, and for the bias have assumed that one is able to describe it with a relative accuracy of $15\%$. Finally, the joint Fisher matrix is the sum of all the Fisher matrices transformed to a common frame. All the calculations done here assume an underlying model with baryonic features in the matter power spectrum.    
Fig. \ref{fig16} presents results for the previously described shallow survey with full sky coverage while the error ellipses in Fig. \ref{fig17} apply to the deep and narrow survey. In both figures dashed lines correspond to the single and solid lines to the multiple bin case. Due to the fact that with a single bin one is able to measure the shape of the power spectrum with a higher precision than in the case of multiple bins, we see from the above figures that in general stronger constraints on $\Omega_{\rm m}$ are obtained. On the other hand constraints on $w_0$ are much stronger in the multiple bin case due to the increased knowledge about the redshift derivatives. We also performed calculations taking two redshift bins and the results were already rather close to the solid curves in the figures above which can be interpreted as an indication that any further redshift slicing would not improve constraints on $w_0$. Also one should not increase the number of redshift bins much above the maximally used values of three and four since then the wavevector bins with width $\Delta k = 0.005\,h\,\mathrm{Mpc}^{-1}$ would become highly correlated and the above Fisher matrix calculation would not be meaningful.

Probably the most interesting constraints in Fig. \ref{fig16} and \ref{fig17} are the ones for $\Omega_{\rm m}$ (or $\Omega_{\rm DE}$ since we have assumed flat models) and $w_0$. Error ellipses in the $\Omega_{\rm m}$-$w_0$ plane are also given in Fig. \ref{fig18} for various different assumptions. The top panel here corresponds to the PLANCK-type and the lower one to the SPT-type of survey. The order of ellipses in the top panel starting from the bottommost one is as follows: (1) clustering signal of the model without baryonic oscillations + CMB priors, (2) (1) + prior on bias, (3) clustering signal of the ``wiggly'' model, (4) (3) + CMB priors, (5) (4) + prior on bias. The only difference in the lower panel is the reversed order of (2) and (3). The inset in the lower plot displays the constraints obtainable (again starting from the bottommost ellipse) with (1) SPT, (2) PLANCK, (3) SPT + PLANCK, (4) the survey with SPT characteristics covering the full sky. For all of the cases shown in the inset we have included CMB and bias priors. The constraint ellipses for the ``clustering only'' case assuming a ``smooth'' model would fill almost all the plot area and for the sake of clarity we have not displayed them here. It is evident from Fig. \ref{fig18} that the model with baryonic oscillations is performing much better compared to its smoothed counterpart. Adding prior information (in contrast to the ``smooth'' case) does not result here in a strong improvement i.e. the clustering signal alone already has a significant constraining power.

Table \ref{tab1} lists the principal components of the clustering analysis only i.e. no CMB and bias priors included. Also we have marginalized over bin bias parameters. This serves as a compact way of summarizing our results. The principal components are given in the form:
\begin{equation}\label{eq34}
\prod \limits_{{\rm j}=1}^{5}\left(\frac{\Theta_{\rm j}}{\Theta^{\rm{fid}}_{\rm j}}\right)^{e_{\rm j}^{\rm{i}}} = 1 \pm \frac{1}{\sqrt{\lambda^{\rm{i}}}} \quad \mathrm{for\ all} \quad i=1\ldots5.
\end{equation}
Here $e_{\rm j}^{\rm{i}}$ is the $j$-th component of the $i$-th eigenvector and $\lambda^{\rm{i}}$ is the $i$-th eigenvalue of the matrix $\widetilde{F_{\rm ij}}=\Theta^{\rm{fid}}_{\rm i}\Theta^{\rm{fid}}_{\rm j}F_{\rm ij}$ (no summation over indices). The parameter vectors $\mathbf{\Theta}=(\Omega_{\rm m},\sigma_8,\Omega_{\rm b},h,w_0)$ and $\mathbf{\Theta^{\rm{fid}}}=(0.27,0.84,0.044,0.71,-1.0)$.

\subsubsection{Comparison to previous work}   
Probably the two closest works to ours are \citet{2004ApJ...613...41M} and \citet{2003PhRvD..68f3004H}. In \citet{2004ApJ...613...41M} the authors discuss constraints obtainable by combining the cluster power spectrum with independent information from cluster number counts. Unfortunately they do not present results separately for the clustering signal only. Moreover, they use an isotropized power spectrum which leads to a significant loss of information, especially when the spectra contain baryonic features. In  \citet{2003PhRvD..68f3004H} the authors use the full 2D power spectrum of galaxy clusters although they do not take into account light-cone effects. Since the light-cone power spectrum is a blend of differently deformed power spectra (if we are away from the reference model point) some loss of baryonic features will result as seen from Fig. \ref{fig15}. Also in their analysis they used $k$-modes up to $0.15$ as opposed to our adopted value of $0.1\,h\,\mathrm{Mpc}^{-1}$. As seen from Fig. \ref{fig5} at $k=0.15\,h\,\mathrm{Mpc}^{-1}$ the cluster power spectrum already differs quite significantly from the simple linear one, particularly for the more massive systems. To simplify the comparison our results with \citet{2003PhRvD..68f3004H}, we used identical CMB priors. An additional difference is that we do not allow the spectral index of the power spectrum to vary but keep it fixed to $n=1$. In total their results for the SPT-type of survey are significantly more optimistic, e.g. the constraints on $w_0$ agree roughly with our SPT+PLANCK case, however, in the case of $\Omega_{\rm m}$ an approximate agreement is achieved with our SPT full sky example.
\section{Conclusions}   
In this paper we studied the clustering of SZ-selected galaxy clusters on a past light-cone with particular emphasis on constraining the properties of DE. We implemented an extended Press-Schechter type of analytical model as described e.g. in \citet{1999MNRAS.308..119S}. The description of the calculation of the light-cone power spectra (e.g. \citealt{1999ApJ...527..488Y}) was modified slightly to incorporate other than simple number weighting schemes. The analytical model was extensively calibrated using the outputs from the VIRGO Consortium's Hubble Volume simulations. With a little bit of fine tuning we were able to match analytical light-cone power spectra with ones extracted from the simulations to an accuracy better than $20\%$. The SZ scaling relations were calibrated so as to get a good match to the number count results from the state-of-the-art hydrodynamical simulations of \citet{2002ApJ...579...16W}. 
Having a well calibrated analytical model we investigated how accurately future SZ surveys like PLANCK and SPT could determine the cluster power spectrum and whether they would be able to detect traces of baryonic oscillations. Also we made use of VIRGO simulation outputs to build cluster catalogs for various survey depths and sensitivity limits. We showed that the aforementioned blank sky SZ surveys will be able to improve the detection of acoustic features based on the SDSS LRG sample. To obtain a high-fidelity detection of the baryonic oscillations, one has to reach a relative accuracy $\sim 5\%$ if the wavenumber bin width $\Delta k=0.005\,h\,\mathrm{Mpc}^{-1}$ is used. This seems hard to achieve with surveys like SPT still having relatively narrow sky coverage. On the other hand, for surveys with a wide sky coverage the prospects seem to be better e.g. with $\sim 25,000$ most massive clusters up to redshift $\sim 0.6$ one should be able to reach a relative accuracy $\sim 4.5\%$ at $k=0.05\,h\,\mathrm{Mpc}^{-1}$ i.e. roughly the scale where one expects to see the first major acoustic feature.

In the last part of the paper we carried out a Fisher matrix forecasting analysis for cosmological parameters, which add up to nine in the case of four redshift bins: $\Omega_{\rm m}$, $\sigma_8$, $\Omega_{\rm b}$, $h$, $w_0$ plus a free bias parameter for each bin. We included prior information for $\Omega_{\rm m}h^2$ and $\Omega_{\rm b}h^2$ from CMB angular power spectrum studies and also constrained the possible values for the bias parameters. A prior on bias parameters only has a significant effect in the case of models with smooth power spectra i.e. models with acoustic oscillations have enough constraining power to give an estimate of bias parameters from the survey itself. The most interesting constraints are obtained for $\Omega_{\rm m}$ and $w_0$. Wide and rather shallow surveys like PLANCK in combination with a CMB prior on $\Omega_{\rm m}h^2$ are able to provide strong constraints on $\Omega_{\rm m}$ or in the case of flat models equivalently on $\Omega_{\rm DE}=1-\Omega_{\rm m}$. The constraints on $w_0$ on the other hand are not as good as the ones obtained by deeper and narrower surveys with the characteristics of SPT due to the lack of higher redshift objects. We also give our results as the principal components of the Fisher matrix that should allow for an easy way of comparison and also for a fast way of incorporating these constraints into further parameter forecasting studies.  

\acknowledgements{I thank Rashid Sunyaev for useful discussions and Maximilian
  Stritzinger for proofreading the paper. I am grateful to the anonymous referee for helpful comments and suggestions. Also I acknowledge the support provided through the European Community's Human Potential Programme under contract HPRN-CT-2002-00124, CMBNET, and the ESF grant 5347. \\ The simulations in this paper were carried out by the Virgo Supercomputing Consortium using computers based at the Computing Centre of the Max-Planck Society in Garching and at the Edinburgh parallel Computing Centre. The data are publicly available at http://www.mpa-garching.mpg.de/NumCos}

\bibliographystyle{aa}

\end{document}